\documentclass[showpacs,preprintnumbers,amsmath,amssymb]{revtex4}
\usepackage{amsmath,amssymb,graphics,epsfig,subfigure}
\usepackage{color}

\begin{document}
\renewcommand{\baselinestretch}{1.3}
\title{Shadow cast by Kerr-like black hole in the presence of plasma in Einstein-bumblebee gravity}

\author{Hui-Min Wang,
        Shao-Wen Wei\footnote{weishw@lzu.edu.cn, corresponding author}}

\affiliation{Lanzhou Center for Theoretical Physics, Key Laboratory of Theoretical Physics of Gansu Province, School of Physical Science and Technology, Lanzhou University, Lanzhou 730000, People's Republic of China,\\
 Institute of Theoretical Physics $\&$ Research Center of Gravitation,
Lanzhou University, Lanzhou 730000, People's Republic of China,\\
 Academy of Plateau Science and Sustainability, Qinghai Normal University,  Xining 810016, P. R. China}

\begin{abstract}
In the Einstein-bumblebee gravity, the Lorentz symmetry is spontaneously broken by a vector field. In this paper, we attempt to test the Lorentz symmetry via the observation of the shadow cast by the Kerr-like black hole with or without plasma. A novel phenomenon of the Lorentz-violating parameter on the shadow is observed. The result shows that when the observer gradually moves from the poles to the equatorial plane, the shadow radius $R_{\rm s}$ firstly decreases and then increases with the Lorentz-violating parameter. Such nonmonotonic behavior provides us an important understanding on the black hole shadow in the Einstein-bumblebee gravity. Besides, three more distortion observables are calculated, and found to increase with the Lorentz-violating parameter. Moreover, when a homogeneous plasma is present, the motion of the photon is analyzed. We further observe that the refractive index shrinks the size, while enhances the deformation of the shadow. Finally, adopting the observed data of the diameter of M87$^*$, we find the refractive index is more favored in (0.914, 1).
\end{abstract}

\keywords{Bumblebee gravity, shadow, plasma}

\pacs{04.50.Kd, 04.25.-g, 04.70.-s}

\maketitle

\section{Introduction}
\label{secIntroduction}

Black hole is one of the fascinating objects in the fields of astrophysics and theoretical physics. Due to the limitation of observation techniques, black holes have never been observed directly before. This situation does not change until April, 2019. It was the Event Horizon Telescope (EHT) who gave the first image of the supermassive black hole in the center of the giant elliptical galaxy M87 \cite{KAkiyama1,KAkiyama2,KAkiyama3,KAkiyama4,KAkiyama5,KAkiyama6}. The image clearly shows the shadow of M87*. As is well known, the formation of such shadow is related to the strong gravity of a black hole. Generally speaking, if a black hole exists between a light source and an observer, photons emitted from the source with small orbital angular momentum will fall into the black hole unavoidably, while these with large orbital angular momentum will escape from the black hole at some turning points and then reach the observer. Accordingly, from the point of view of an observer, these ``missing'' photons will create a two-dimensional dark zone in the sky, which is called the black hole shadow.

The trajectory of photons is determined by the geometry of the spacetime around a black hole. Thus, the shadow closely depends on the nature of the black hole. It provides an opportunity for testing the black hole via the shadow. The shadow cast by a non-rotating Schwarzschild black hole was first studied by Synge and Luminet~\cite{JLSynge1,JPLuminet}, and the rotating Kerr black hole shadow was investigated by Bardeen~\cite{JMBardeen}. In order to study black hole shadow systematically and intuitively, many astronomical observations were constructed~\cite{CBambi1,KHioki,CBambi2,ZLLi,NTsukamoto,TJohannsen,AAbdujabbarov1,
MGhasemi,RKumar}. Using them, the shadow cast by different types black holes and wormholes has been well studied~\cite{SWWei1,SAbdolrahimi,
Aovgun1,HMWang,SWWei2,SWWei3,MZWang,TZhu,CLiu1,CLiu2,SWWei4,
MKhodadi}.

General speaking, black holes in our universe are more likely to be surrounded by matter, so it is worth to consider the non-vacuum environment. Since in the present universe, the most common phase of ordinary matter is the plasma, it is likely to be the matter around a black hole. On the other hand, plasmas in the universe can not only generate magnetic fields, but also interact with magnetic fields. Usually, they are hot enough to emit electromagnetic radiation in a large range of the electromagnetic spectrum, such as X-rays, radio waves, and gamma rays. Astrophysical plasmas exist in the accretion disk of compact objects~\cite{Ichimaru,Narayan,Hollywood}, and they are also related to the ejected matter in the jet, like the jet of the active galaxy M$87$~\cite{RCThomson,CSReynolds,YYKovalev}. Besides, these plasmas around the black hole act as a kind of medium, and they will inevitably affect the propagation trajectory of photons~\cite{ABroderick}. Therefore, the shape of the black hole shadow will change. For instance, the influence of plasmas on the rotating and non-rotating black hole shadows were studied in Refs.~\cite{VPerlick1,FAtamurotov,AAbdujabbarov2,CQLiu,VPerlick2,HYan,SDastan,ASaha,ADas,GZBabar,MFathi}.

Lorentz invariance is an important principle of the general relativity (GR). However, it is not an exact symmetry at all energies \cite{Mattingly}. When spacetime is discrete, the Lorentz invariance is also no longer applicable. In some other gravity theories, such as the loop quantum gravity, the Lorentz symmetry breaks \cite{CRovelli,AAshtekar1,AAshtekar2,AAshtekar3,AAshtekar4}. Therefore, the test of the Lorentz violation is a challenge to our fundamental physics. Bumblebee gravity, as a simple model of Lorentz violation, was first proposed by Kosteleck\'{y} and Samuel \cite{VAKostelecky1,VAKostelecky2,VAKostelecky3}. When a vector field with a nonzero vacuum expectation value, the Lorentz symmetry is broken. Later some other properties were inviestigated \cite{VAKostelecky4,VAKostelecky5,RBluhm1,VAKostelecky6,RBluhm}. In 2018, Casana {\it et al.} gave the exact Schwarzschild-like solution in a Einstein-bumblebee gravity model \cite{RCasana}. Shortly thereafter, many groups have studied the thermodynamics and dynamics of black holes in the Einstein-bumblebee gravity \cite{CLiu1,DAGomes,SKanzi,CDing1,ZLi,AAli,SChen,RVMaluf,SKJha1,
CDing2,Carvalho,SKJha2}.

Therefore, it is interesting to explore the behavior of the null geodesics of the black hole in the Einstein-bumblebee gravity. In this paper, we start with the exact Kerr-like solution \cite{CDing1}. Then after examining the equations of motion for the photon, we study the shadow cast by the rotating and non-rotating black holes. The influence of the Lorentz violation on the size and deformation of the shadow are investigated. The radius and distortion observables are calculated. Furthermore, when a homogeneous plasma is present, we also examine the shadow shape in details.

This paper is organized as follows. In Sec.~\ref{theory}, we briefly review the black hole solution in the Einstein-bumblebee gravity. In Sec.~\ref{shadow}, employing the null geodesics, we study the black hole shadow. Different observables are calculated. The influence of the Lorentz violation on the shadow is also discussed. When a homogeneous plasma is present, we examine the  effects of plasma on photon trajectory, black hole shadow, as well as the deflection angle of light in Sec.~\ref{plasma}. In Sec.~\ref{constraint}, we place the constraints on the parameters of Lorentz violation and the refractive index of the plasma via the observations of M87*. Finally, we summarize and discuss our results. In this paper, we adopt the geometric units $\hbar=G=c=1$.

\section{Einstein-bumblebee gravity and the Kerr-like solution}
\label{theory}

In this context we will consider the null geodesics around a Kerr-like black hole in the Einstein-bumblebee gravity, where the Lorentz symmetry is broken by introducing a extra vector field called the bumblebee field $B_{\mu}$.

\begin{figure*}
\begin{center}
\includegraphics[width=8cm]{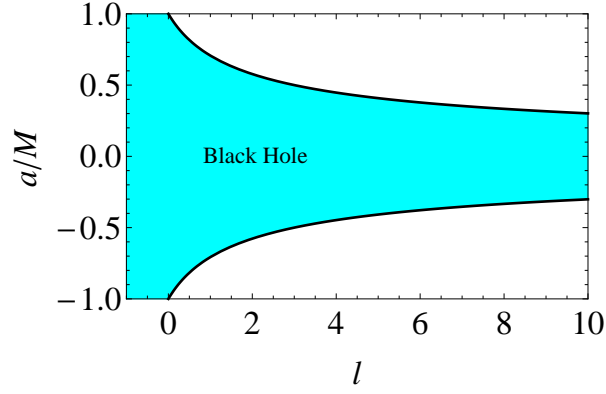}
\caption{The blue region represents the allowable range of black hole.}\label{horizon}
\end{center}
\end{figure*}

This field acquires a nonvanishing vacuum expectation value and induces a spontaneous Lorentz symmetry breaking under a suitable potential. The action of the theory is~\cite{RBluhm1,VAKostelecky1,RCasana}
\begin{eqnarray}
S_B=\int d^{4}x \sqrt{-g}(\mathcal{L}_{g}+\mathcal{L}_{gB}
+\mathcal{L}_{\text{K}}+\mathcal{L}_{V}
+\mathcal{L}_{\text{M}}),\label{eq:action}
\end{eqnarray}
with
\begin{subequations}
\begin{eqnarray}
\mathcal{L}_{g}&=&\frac{1}{16\pi}R, \\
\mathcal{L}_{gB}&=&\frac{1}{16\pi}\varrho B^{\mu}B^{\nu}R_{\mu\nu},\label{nonminimalc1} \\
\mathcal{L}_{\text{K}}&=&-\frac{1}{4}B^{\mu\nu}B_{\mu\nu}, \\
\mathcal{L}_{V}&=&-V(B_{\mu}B^{\mu}),
\end{eqnarray}
\end{subequations}
where $\mathcal{L}_{g}$,  $\mathcal{L}_{\text{K}}$, $\mathcal{L}_{gB}$, $\mathcal{L}_{V}$, and $\mathcal{L}_{\text{M}}$ are the usual Einstein-Hilbert term, kinetic term of the bumblebee field, coupling term between the bumblebee field and gravity, potential term, and matter Lagrangian, respectively. From the coupling term~\eqref{nonminimalc1}, we know that the bumblebee field is non-minimally coupling to the gravity. In the description of the gravity-bumblebee coupling term, $\varrho$ acts as a real coupling constant (with mass dimension $-1$). The corresponding bumblebee field strength and the potential are defined as
\begin{eqnarray}
B_{\mu\nu}=\partial_{\mu}B_{\nu}-\partial_{\nu}B_{\mu},\nonumber \\
V=V(B_{\mu}B^{\mu}\pm b^2),
\end{eqnarray}
where $b^2$ is a real positive constant. The gravitational field equation corresponding to the action \eqref{eq:action} reads
\begin{eqnarray}
R_{\mu\nu}-\frac{1}{2}g_{\mu\nu}R=8\pi T_{\mu\nu},\label{soli}
\end{eqnarray}
in which $T_{\mu\nu}=T_{\mu\nu}^{\text{M}}+T_{\mu\nu}^B$. The Einstein-bumblebee energy momentum tensor $T_{\mu\nu}^B$ is
\begin{eqnarray}\label{momentum}
T_{\mu\nu}^B&=&B_{\mu\alpha}B^{\alpha}_{\;\nu}-\frac{1}{4}g_{\mu\nu} B^{\alpha\beta}B_{\alpha\beta}- g_{\mu\nu}V+
2B_{\mu}B_{\nu}V'\nonumber\\
&~&+\frac{\varrho}{8\pi}\Big(\frac{1}{2}g_{\mu\nu}B^{\alpha}B^{\beta}R_{\alpha\beta}
-B_{\mu}B^{\alpha}R_{\alpha\nu}-B_{\nu}B^{\alpha}R_{\alpha\mu}\nonumber\\
&~&+\frac{1}{2}\nabla_{\alpha}\nabla_{\mu}(B^{\alpha}B_{\nu})
+\frac{1}{2}\nabla_{\alpha}\nabla_{\nu}(B^{\alpha}B_{\mu})
-\frac{1}{2}\nabla^2(B^{\mu}B_{\nu})-\frac{1}{2}
g_{\mu\nu}\nabla_{\alpha}\nabla_{\beta}(B^{\alpha}B^{\beta})\Big).
\end{eqnarray}
Solving the field equation (\ref{soli}) with $T_{\mu\nu}^{\text{M}}=0$, the exact rotating Kerr-like black hole solution was obtained \cite{CDing1}
\begin{eqnarray}\label{metric}
ds^2=- \Big(1-\frac{2Mr}{\rho^2}\Big)dt^2-\frac{4Mra\sqrt{1+l}\sin^2\vartheta}{\rho^2}
dtd\phi+\frac{\rho^2}{\Delta}dr^2+\rho^2d\vartheta^2
+\frac{A\sin^2\vartheta}{\rho^2} d\phi^2,
\end{eqnarray}
where
\begin{eqnarray}
\rho^2&=&r^2+(1+l)a^2\cos^2\vartheta,\nonumber\\ \Delta&=&\frac{r^2-2Mr}{1+l}+a^2,\nonumber\\
A&=&\big[r^2+(1+l)a^2\big]^2-\Delta(1+l)^2 a^2\sin^2\vartheta.
\end{eqnarray}
The parameter $l$ is related to the spontaneous Lorentz symmetry breaking of the vacuum of the Einstein-bumblebee vector field $B_{\mu}$. The Kerr solution in GR will be recovered when the LV parameter $l$ tends to zero. By solving $\Delta=0$, the radii of the black hole horizons can be obtained as follows
\begin{eqnarray}
r_{\pm}=M\pm\sqrt{M^2-a^2(1+l)}.
\end{eqnarray}
It is obvious that the mass and spin parameter for a black hole should satisfy
\begin{eqnarray}
\frac{|a|}{M}\leq \frac{1}{\sqrt{1+l}},
\end{eqnarray}
where the inequality is saturated for the extremal black holes. We show the above constraint in the parameter range in Fig.~\ref{horizon}. The blue region is for the black holes while other regions are for the naked singularities. As one can see, the dimensionless black hole spin decreases when the Lorentz-violation becomes stronger. Especially, a negative LV parameter allows $|a|/M>1$, which is quite different from the case in GR. When the LV parameter $l$ takes zero, the maximum of $|a|/M$ decreases to $1$, which indicates that the black hole spin given in this theory is consistent with that in GR when the LV parameter is zero. Here we will adopt the mass $M$ as the unit.

\section{Black hole shadow}
\label{shadow}
We know that shadows are one of the most intuitive evidence of the existence of black hole. Especially the release of the image provides a stronger evidence for the existence of black hole. Therefore, in this section, we would like to examine the shadow cast by the Kerr-like black holes in the Einstein-bumblebee gravity.

\subsection{Photon orbit}
Here we shall study the equations of motion of the photon orbiting the black hole. The motion of the photon in the background of black holes is given by the geodesics. By solving the geodesic equation, we can study the motion of the photon. However, considering that the process of directly solving the geodesics is very difficult, we turn to another way, namely, the Hamilton-Jacobi approach. The Hamilton-Jacobi equation in the Einstein-bumblebee gravity spacetime with the inverse metric tensor $g^{\mu\nu}$ is given by
\begin{equation}
-\frac{\partial S}{\partial\lambda}=\frac{1}{2}g^{\mu\nu}\frac{\partial S}{\partial x^\mu}\frac{\partial S}{\partial x^\nu},\label{HJ}
\end{equation}
where $\lambda$ is the affine parameter. The Jacobi action $S$ can be written as the following formula
\begin{equation}
S=\frac{1}{2}m^2\lambda-Et+L_{\rm z}\phi+S_{r}(r)+S_\vartheta(\vartheta),\label{actionn}
\end{equation}
with two motion constants $E$ and $L_{\rm z}$ are, respectively, the energy and orbital angular momentum in the direction of the axis of symmetry. Besides, $S_{r}(r)$ and $S_\vartheta(\vartheta)$ are the separated parts of the Jacobi action that only depend on $r$ and $\vartheta$, respectively. Substituting \eqref{actionn} into \eqref{HJ}, one can obtain the equations of motion of the photon with $m^2=0$ in the background of a Kerr-like black hole
\begin{eqnarray}
(1+l)\rho^2\frac{dt}{d\lambda}&=&aL_{\rm z}(1+l)-a^2(1+l)^{\frac{3}{2}}E\sin^2\vartheta + \frac{r^2+(1+l)a^2}{\Delta} \left((r^2+(1+l)a^2)E-aL_{\rm z}\sqrt{1+l} \right) ,\label{HS01}\\
 (1+l)\rho^2\frac{d\phi}{d\lambda}&=&\frac{(1+l)L_{\rm z}}{\sin^2\vartheta}
 -a(1+l)^{\frac{3}{2}}E+\frac{a\sqrt{1+l}}{\Delta} \left((r^2+(1+l)a^2)E-aL_{\rm z}\sqrt{1+l} \right),\label{HS02}\\
 \rho^2\frac{dr}{d\lambda}&=&\sqrt{{\cal R}(r)},\label{HS03}\\
 \rho^2\frac{d\vartheta}{d\lambda}&=&\sqrt{\Theta(\vartheta)},\label{HS04}
 \end{eqnarray}
where ${\cal R}(r)$ and $\Theta(\vartheta)$ are given by
\begin{eqnarray}
{\cal R}(r)&=&-\Delta \left({\cal K}+L_{\rm z}^2+a^2E^2 (1+l)-2aL_{\rm z} E\sqrt{1+l} \right)
+ \left(\frac{(r^2+a^2(1+l))E}{\sqrt{1+l}}-aL_{\rm z}\right)^2,\\
\Theta(\vartheta)&=&{\cal K}+(1+l)a^2E^2 \cos^2 \vartheta-L_{\rm z}^2 \cot^2 \vartheta.
\end{eqnarray}
The Carter constant ${\cal K}$ is another constant of the geodesics \cite{BCarter}. Combining with the energy $E$ and the orbital angular momentum $L_{\rm z}$, these three conserved quantities can be used to uniquely determine the orbits of photon in the Kerr-like spacetime. As we mentioned in the Sec.~\ref{secIntroduction}, photons with sufficiently large orbital angular momentum can escape from the black hole, while those with small angular momentum will fall into the black hole. Now we focus on the critical case, the circular null orbit, where the photons will neither escape from or fall into the black hole. The radial motion \eqref{HS03} can also be expressed as
\begin{equation}
 \left( \rho^2\frac{dr}{d\lambda}\right)^2+V_{\rm eff} =0.\label{Veff}
\end{equation}
After defining two new conserved parameters $\xi=L_{\rm z}/E$ and $\eta={\cal K}/E^2$, the effective potential reads
 \begin{equation}
\frac{V_{\rm eff}}{E^2}=\left(\frac{r^2-2Mr}{1+l}+a^2 \right)
\left(\eta+(\xi-a\sqrt{1+l})^2\right)-\frac{\left(r^2+(1+l)a^2 \right)^2}{1+l}+\frac{2a\xi\left(r^2+(1+l)a^2 \right)}{\sqrt{1+l}}-a^2 \xi^2.
\end{equation}
For the unstable circular null orbit, the effective potential satisfies
\begin{eqnarray}
V_{\rm eff}=0,\quad \frac{\partial V_{\rm eff}}{\partial r}=0, \quad \text{and} \quad \frac{\partial^2 V_{\rm eff}}{\partial r^2}<0.
\end{eqnarray}
Solving them, we have
\begin{eqnarray}
\xi&=&\frac{M\left( a^2(1+l)-r^2 \right)+\Delta(1+l)r}{a\sqrt{1+l}(M-r)},\label{yxieta}\\
\eta&=&\frac{r^3 \left( 2a^2(1+l)M-r(5M^2-4Mr+r^2)+2 \Delta(1+l)M \right) }{ a^2(1+l)(M-r)^2}.\label{xieta}
\end{eqnarray}
Then the motion of the photon will be uniquely determined by the above two conserved parameters. The black hole shadow is also closely related with them. In the following sections, we aim to study the property of the shadow.

\subsection{Observables}
\label{observable}

Although these two conserved parameters $\xi$ and $\eta$ determine the motion of the photon, they cannot intuitively describe the shadow seen by a distant observer. In order to specifically portray the shadow, we adopt the celestial coordinates $x$ and $y$
\begin{eqnarray}
x&=&-\xi\csc\theta,\nonumber\\
y&=&\pm\sqrt{\eta+a^2\cos^2\theta-\xi^2\cot^2\theta},\label{coordinate}
\end{eqnarray}
where $\theta$ denotes the angle between the rotation axis of the black hole and the line of sight of the observer. Combining with \eqref{yxieta}and \eqref{xieta}, the shadow can be clearly visualized in $x$-$y$ plane. For certain spin parameter $a$, LV parameter $l$ and observation angle $\theta$, the shadow images has been given in Ref.~\cite{CDing1}.

\begin{figure*}
\begin{center}
\subfigure[]{\label{radius1}
\includegraphics[width=5.5cm]{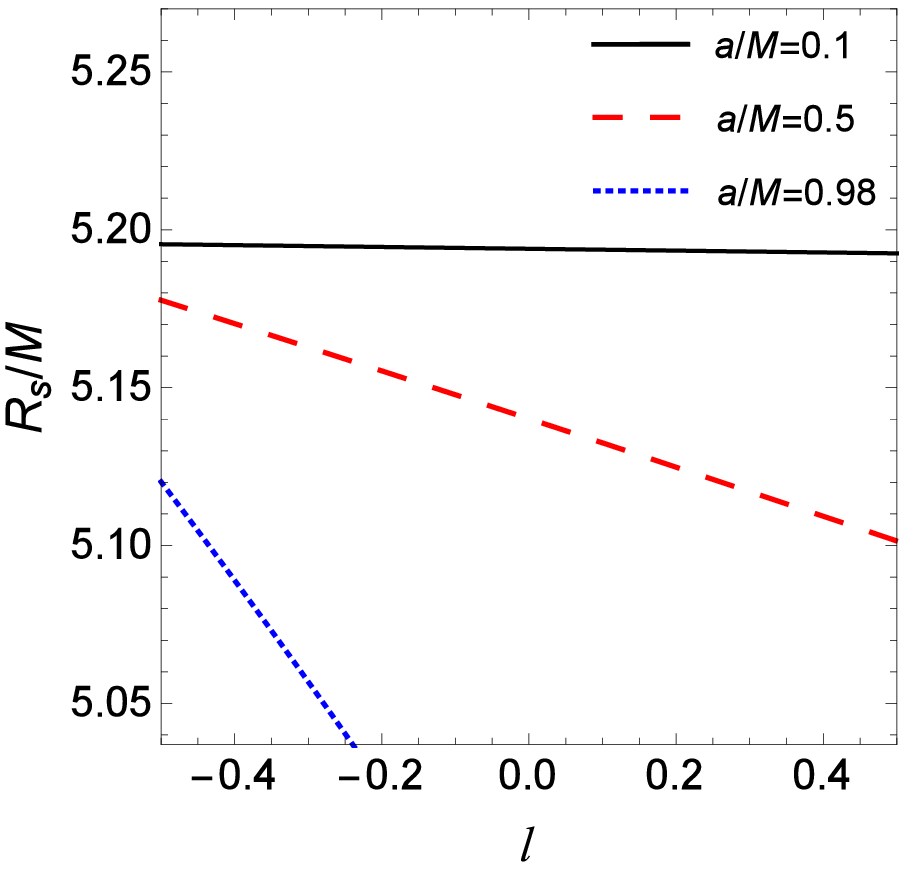}}
\subfigure[]{\label{radius2}
\includegraphics[width=5.5cm]{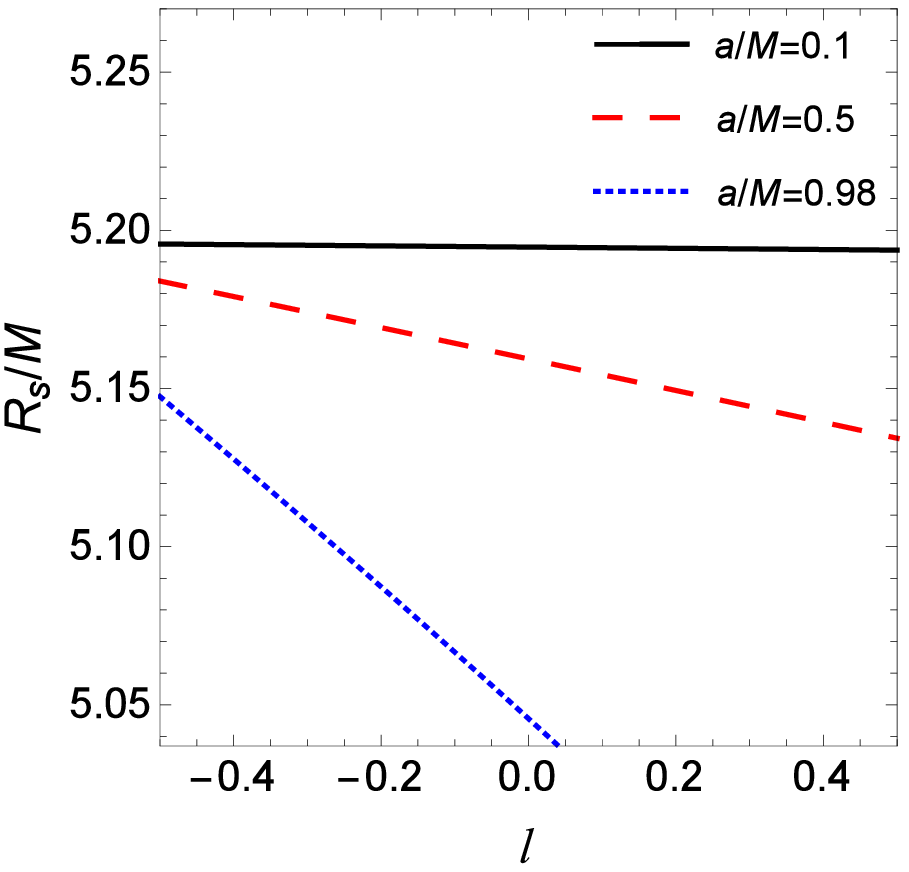}}
\subfigure[]{\label{radius3}
\includegraphics[width=5.5cm]{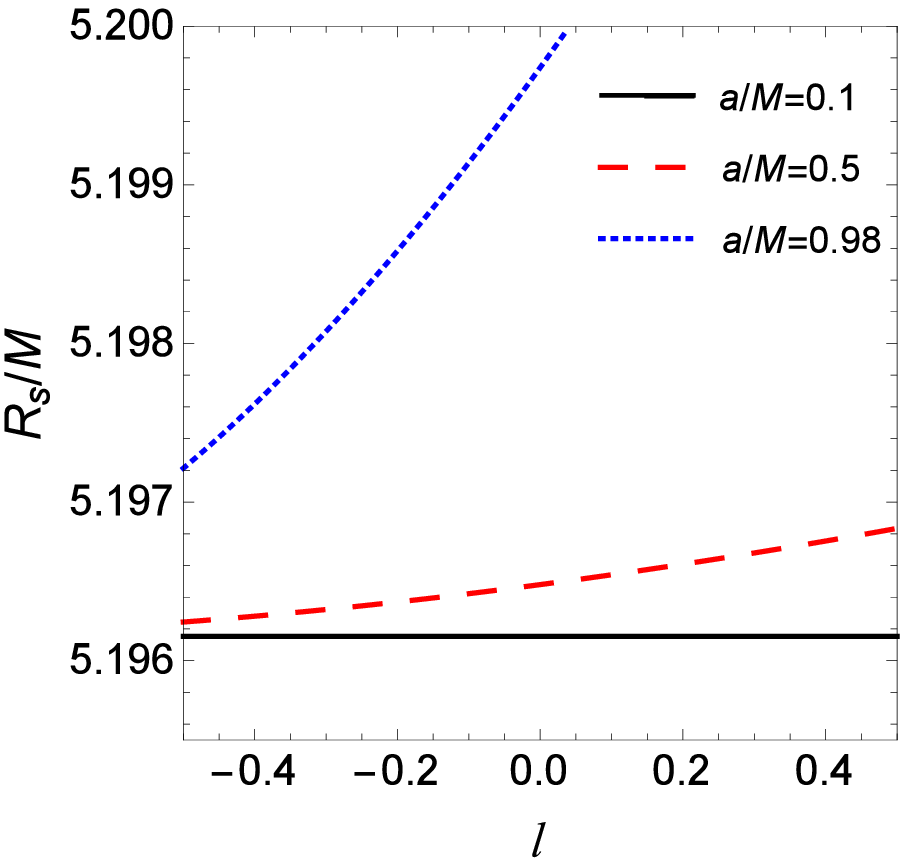}}
\end{center}
\caption{The radius $R_{\rm s}$ of the shadow against the LV parameter $l$. (a)~$\theta=\frac{\pi}{6}$. (b)~$\theta=\frac{\pi}{4}$. (c)~$\theta=\frac{\pi}{2}$.}\label{radiusfig}
\end{figure*}

\begin{figure*}
\begin{center}
\subfigure[]{\label{theta1}
\includegraphics[width=5.5cm]{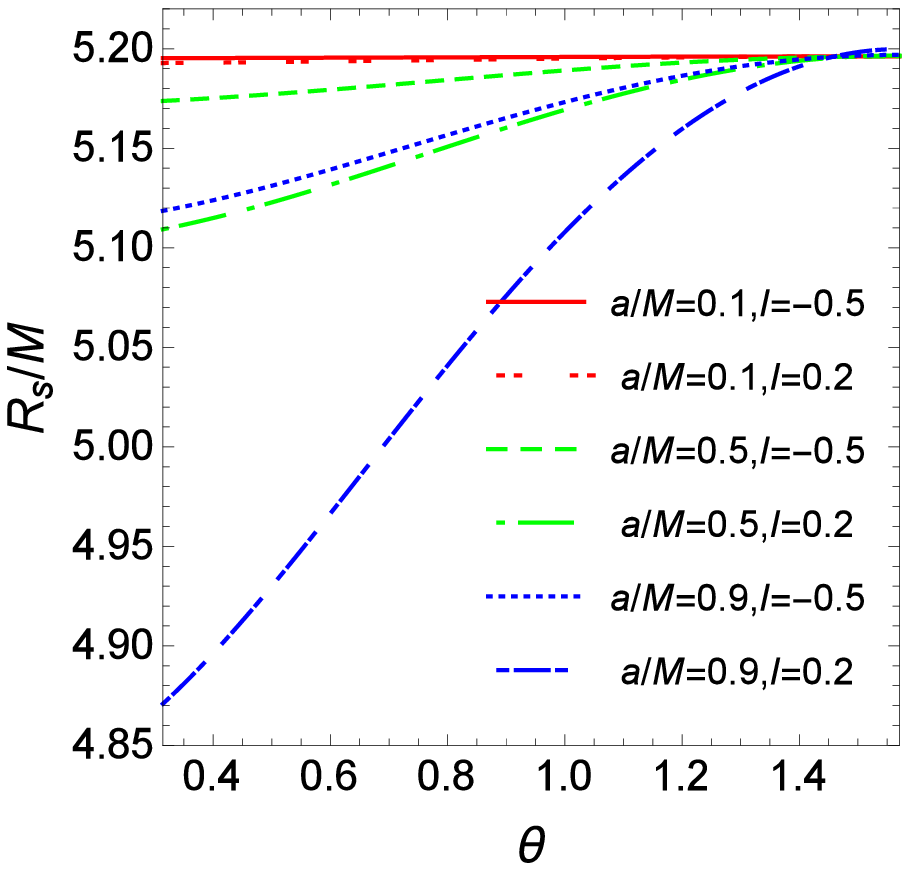}}~~~~~~~~~~
\subfigure[]{\label{theta2}
\includegraphics[width=5.5cm]{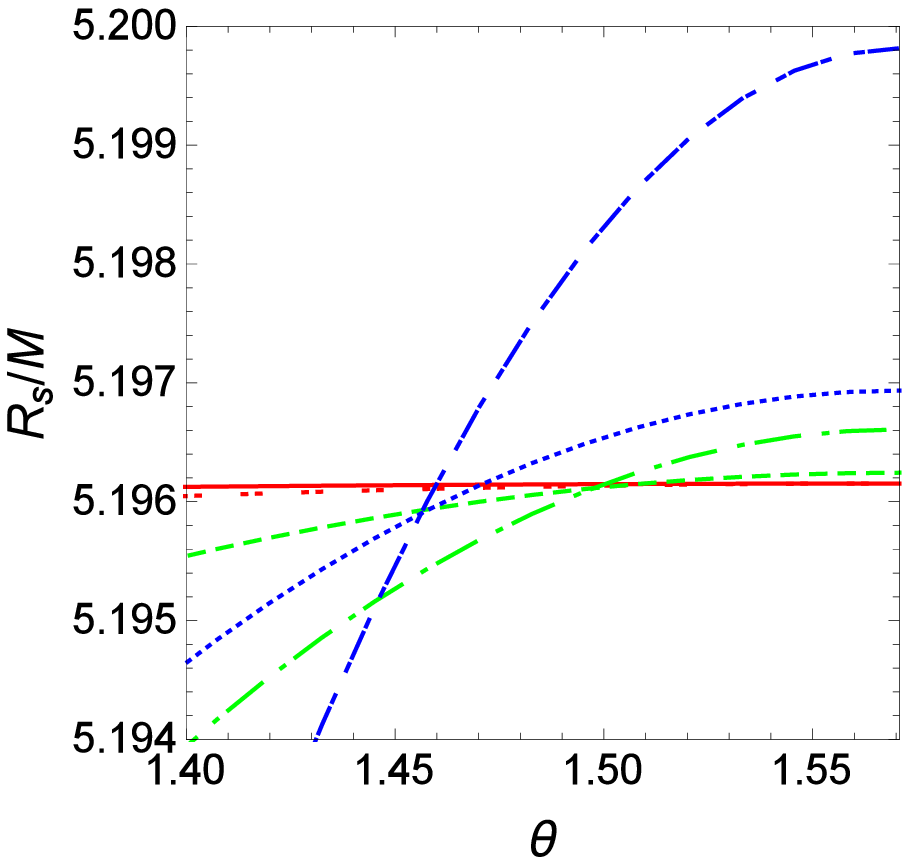}}
\end{center}
\caption{The radius $R_{\rm s}$ of the shadow against the observation angle $\theta$. (a)~The range of $\theta$ is
($\pi/10,\pi/2$). (b)~$\theta$ is near $\pi/2$.}\label{thetafig}
\end{figure*}

To fit the astronomical observation by the theoretical model, observables of the shadow are necessary. For the purpose, various observables are proposed \cite{CBambi1,KHioki,CBambi2,ZLLi,NTsukamoto,TJohannsen,AAbdujabbarov1,
MGhasemi,RKumar}. For a non-rotating black hole, its shadow is a perfect circle in the eyes of a distant observer. However, when the black hole spin is included, the shadow is deformed. In order to measures the shadow size, we adopt the radius $R_{\rm s}$ of the reference circle introduced by Hioki and Maeda \cite{KHioki}. The reference circle is defined as the one passing through the following three points: the top point $(x_{\rm t}, y_{\rm t})$, the bottom point $(x_{\rm b},y_{\rm b})$, and the right point $(x_{\rm r},0)$ of the shadow. After a simple algebraic calculation, $R_{\rm s}$ can be obtained via
\begin{eqnarray}
 R_{\rm s}=\frac{(x_{\rm t}-x_{\rm r})^{2}+y_{\rm t}^{2}}{2(x_{\rm r}-x_{\rm t})}. \label{radius}
\end{eqnarray}
For given spin, we show the radius $R_{\rm s}$ as a function of the LV parameter $l$ for $\theta$=$\frac{\pi}{6}$, $\frac{\pi}{4}$, and $\frac{\pi}{2}$ in Fig.~\ref{radiusfig}. In Figs. \ref{radius1} and \ref{radius2}, we observe a similar pattern for $\theta$=$\frac{\pi}{6}$ and $\frac{\pi}{4}$. For a fixed spin $a$, the radius $R_{\rm s}$ decreases with $l$. However, when $\theta$=$\frac{\pi}{2}$, the result reverses, and $R_{\rm s}$ increases with $l$. This is a novel phenomenon in Einstein-bumblebee gravity dominated by the observation angle $\theta$, which is quite different from the Kerr black hole in GR with $l$=0. In order to show it, we describe $R_{\rm s}$ as a function of the observation angle $\theta$ for certain values of $a$ while with $l$=-0.5 and 0.2 in Fig. \ref{thetafig}. From Fig.~\ref{theta1}, we find that the radius of the black hole shadow increases with the observation angle $\theta$. For an observer far away from the equatorial plane, the shadow of the fast-rotating black hole is smaller than that of a slower one with the same $l$. However, this gap narrows with the observation angle, and eventually all the curves tend to overlapped when $\theta$ is approaching $\pi/2$. In order to show it more clearly, we zoomed in in Fig.~\ref{theta2}. It tells that when $\theta$ increases to a certain value, $R_{\rm s}$ corresponding to the larger $a$ will exceed the smaller one, resulting in the reverse situation that the faster the black hole rotates, the larger the shadow is.

\begin{figure*}
\begin{center}
\subfigure[]{\label{delta1}
\includegraphics[width=5.5cm]{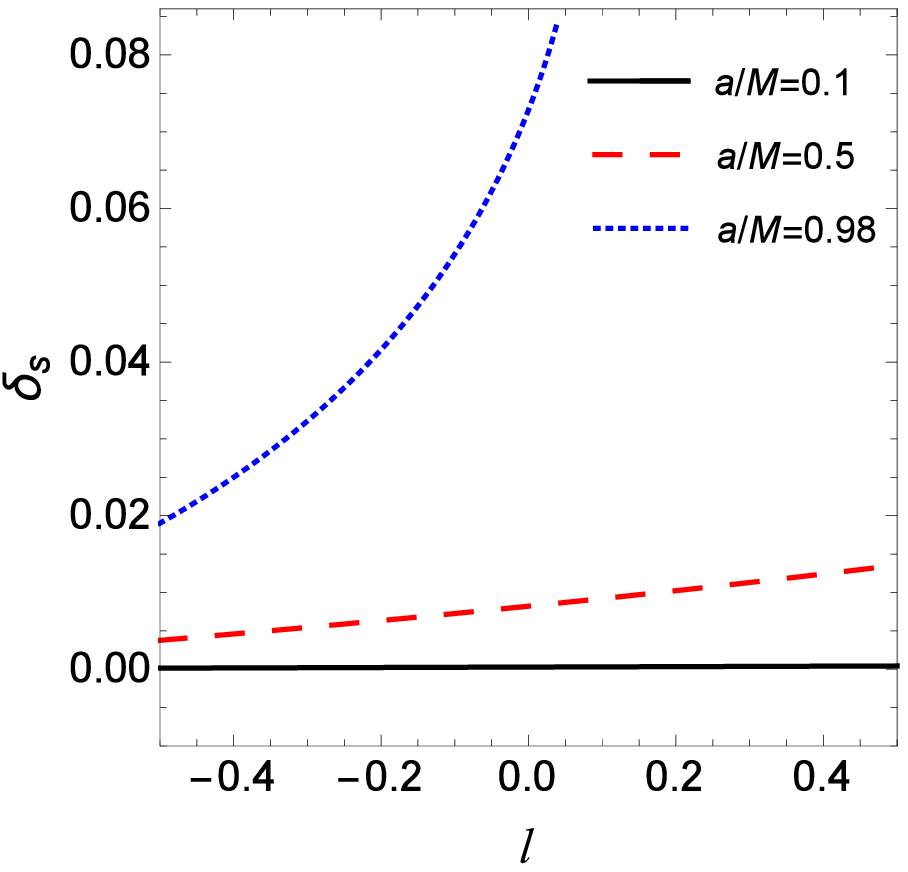}}
\subfigure[]{\label{delta2}
\includegraphics[width=5.5cm]{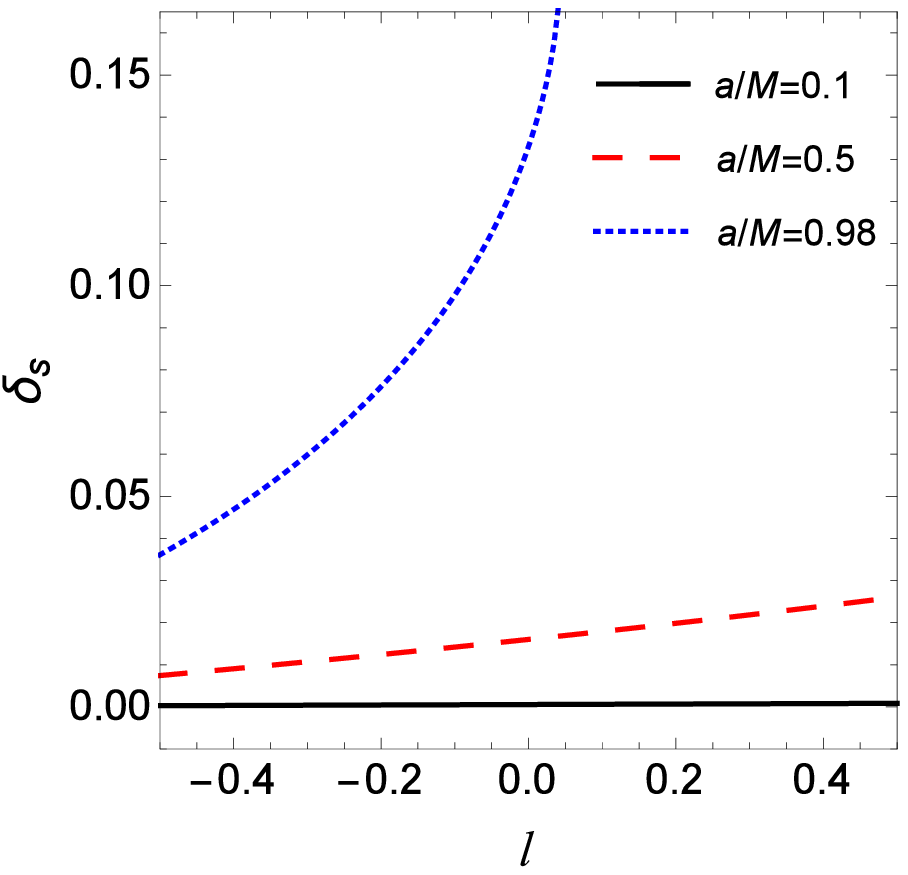}}
\subfigure[]{\label{delta3}
\includegraphics[width=5.5cm]{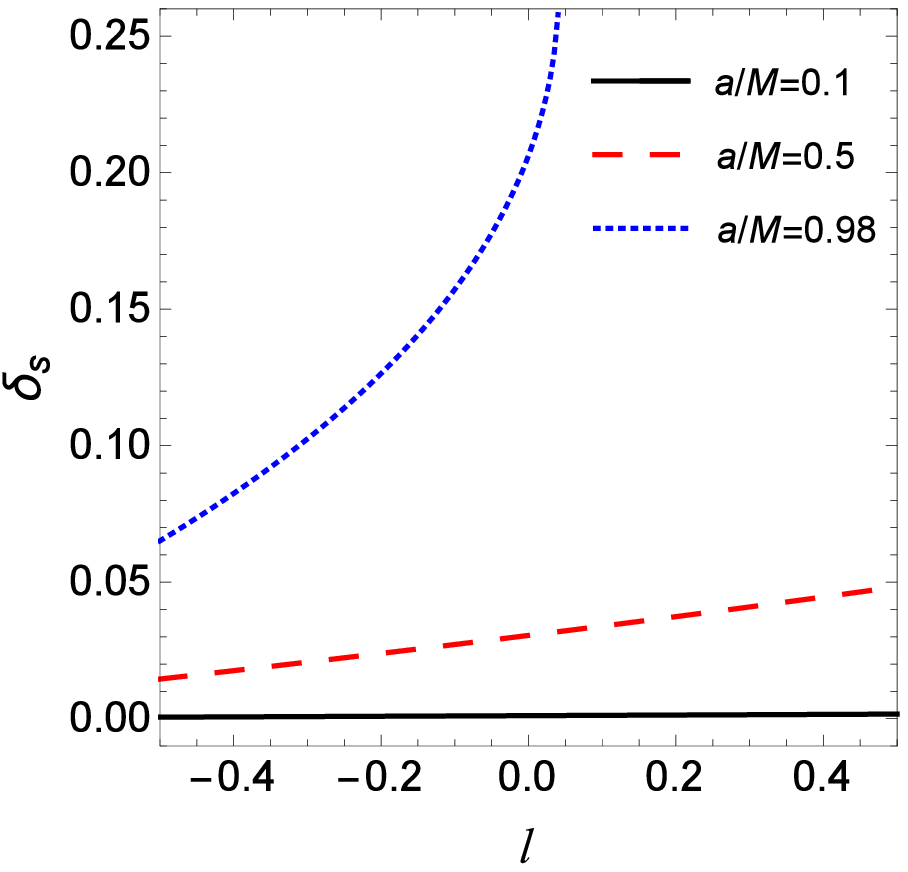}}
\end{center}
\caption{The distortion parameter $\delta_{\rm s}$ of the shadow. (a)~$\theta=\frac{\pi}{6}$. (b)~$\theta=\frac{\pi}{4}$. (c)~$\theta=\frac{\pi}{2}$.}\label{deltafig}
\end{figure*}

As mentioned above, the shadow will be deformed for a rotating black hole. In order to describe it, Hioki and Maeda constructed another observable, the distortion parameter $\delta_{\rm s}$~\cite{KHioki}, which describes the degree of deformation between the shadow and the reference circle. It can be expressed as
\begin{eqnarray}
 \delta_{\rm s}=\frac{D}{R_{\rm s}},
\end{eqnarray}
with $D$ denoting the difference between two characteristic points on the same side, i.e., one is the leftmost point of the shadow $(x_{\rm l}, 0)$ and the other is the endpoint of the reference circle $(\tilde{x}_{\rm l}, 0)$. The behavior of $\delta_{\rm s}$ is exhibited in Fig.~\ref{deltafig} as function of $l$. It is easy to see that for different $a$, $\delta_{\rm s}$ increases monotonically with $l$. This result is also independent of $\theta$. Another result we observe is that, for small spin $a$, the deformation of the shadow is almost imperceptible. For examples, when $l=-0.5$ and $a=0.1$, $\delta_{\rm s}=0.0139\%, 0.0279\%$ and $0.0557\%$ for $\theta=\pi/6, \pi/4$, and $\pi/2$, respectively. However for large $a$, we can find from Fig.~\ref{deltafig} that $\delta_{\rm s}$ grows quickly with $l$. Besides, with the observer approaching the equatorial plane gradually, the deformation of the shadow becomes more obvious. For example, when $a=0.98$ and $l=0.04$, $\delta_{\rm s}=8.4567\%$ at $\theta=\pi/6$ and the deformation increases to 25.7879$\%$ at $\theta=\pi/2$.

\begin{figure*}
\begin{center}
\subfigure[]{\label{epsilon1}
\includegraphics[width=5.5cm]{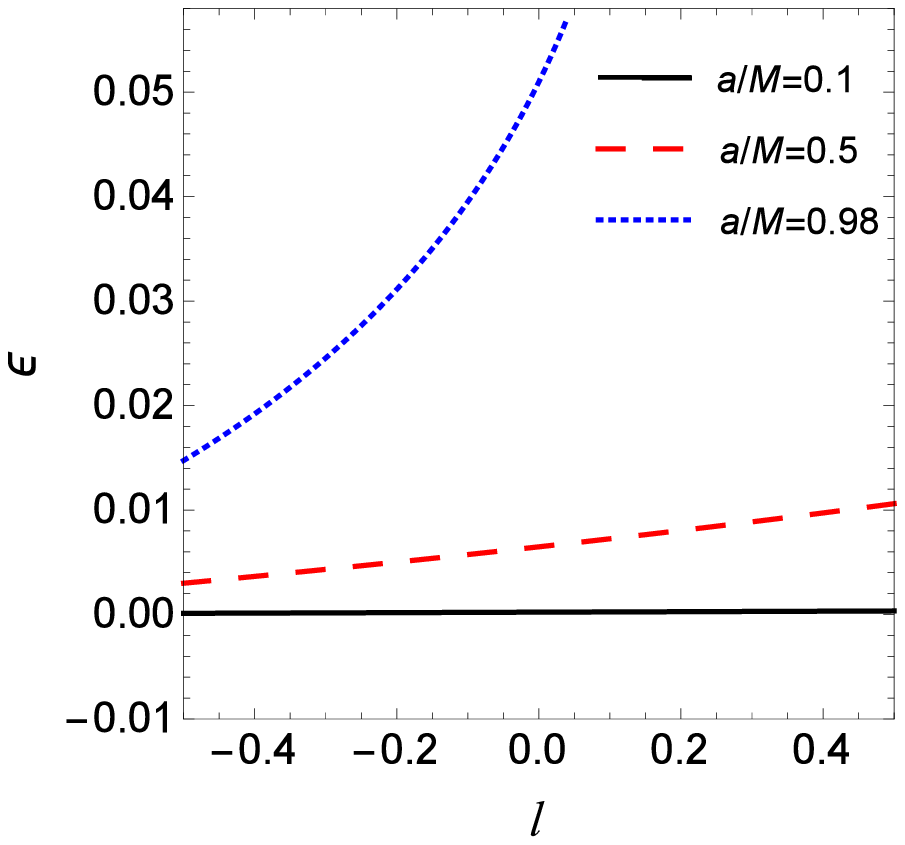}}
\subfigure[]{\label{epsilon2}
\includegraphics[width=5.5cm]{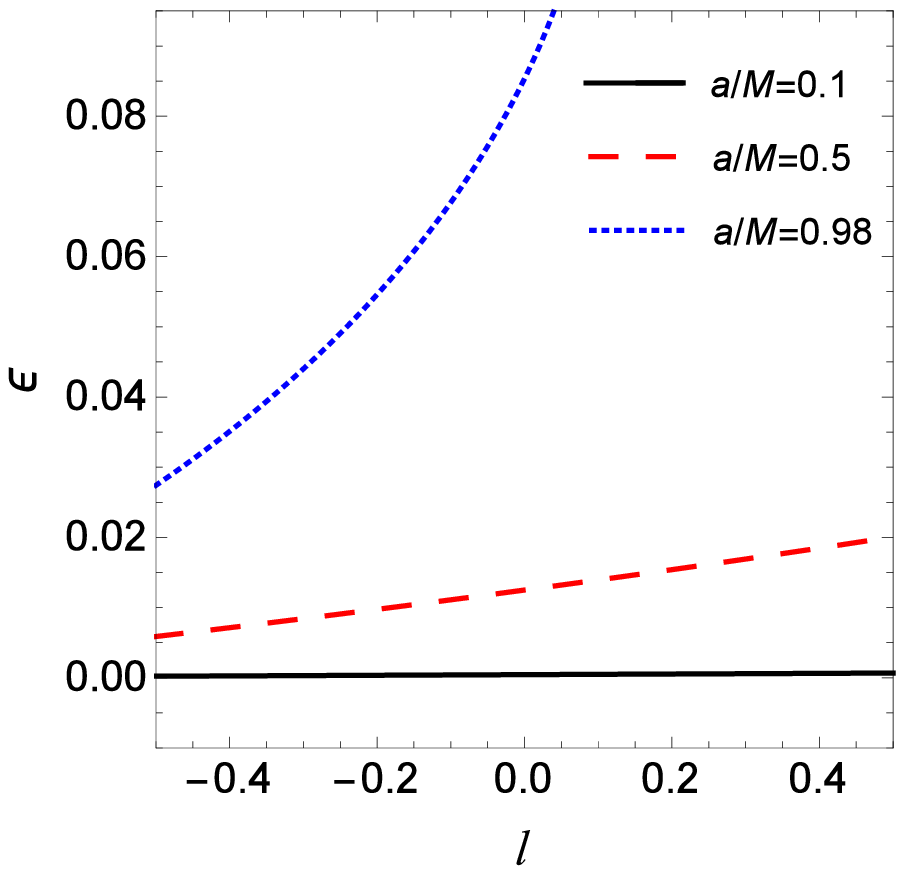}}
\subfigure[]{\label{epsilon3}
\includegraphics[width=5.5cm]{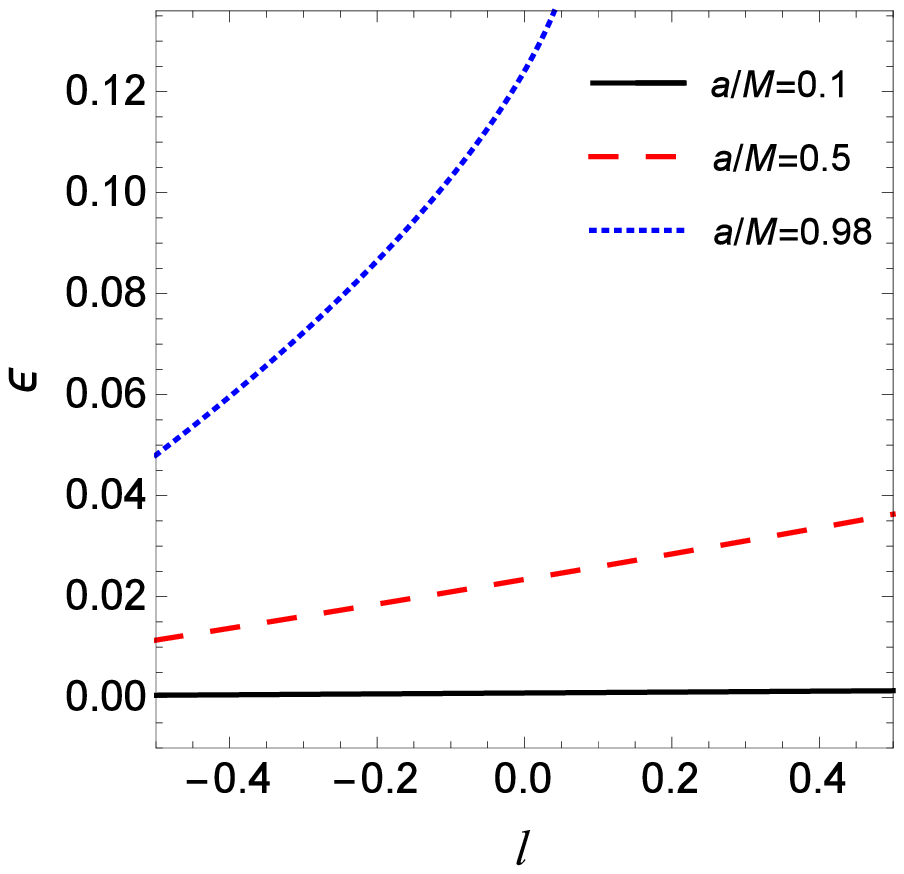}}
\end{center}
\caption{The distortion parameter $\epsilon$ of the shadow. (a)~$\theta=\frac{\pi}{6}$. (b)~$\theta=\frac{\pi}{4}$. (c)~$\theta=\frac{\pi}{2}$.}\label{epsilonfig}
\end{figure*}

\begin{figure*}
\begin{center}
\subfigure[]{\label{ks1}
\includegraphics[width=5.5cm]{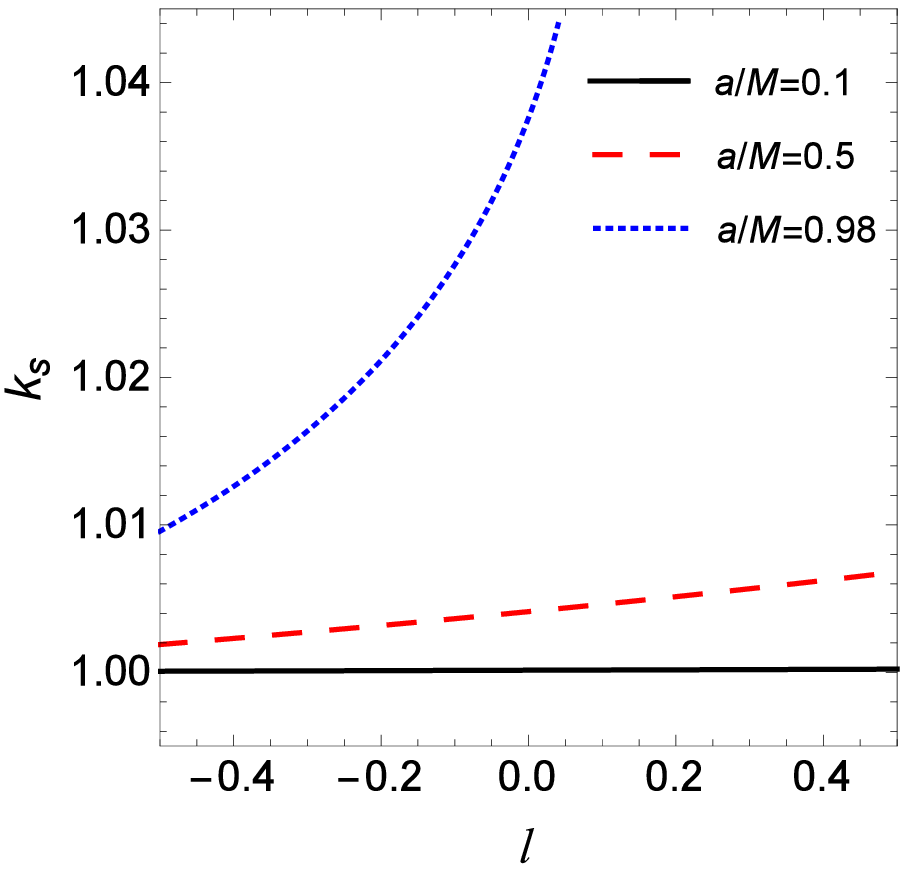}}
\subfigure[]{\label{ks2}
\includegraphics[width=5.5cm]{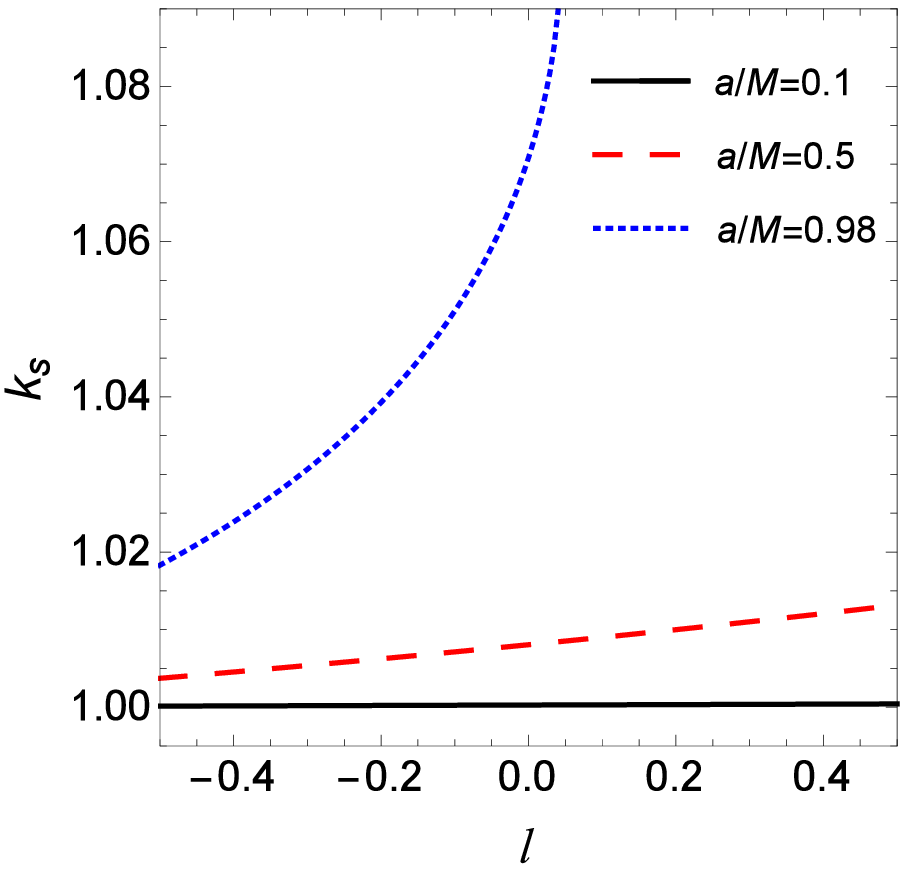}}
\subfigure[]{\label{ks3}
\includegraphics[width=5.5cm]{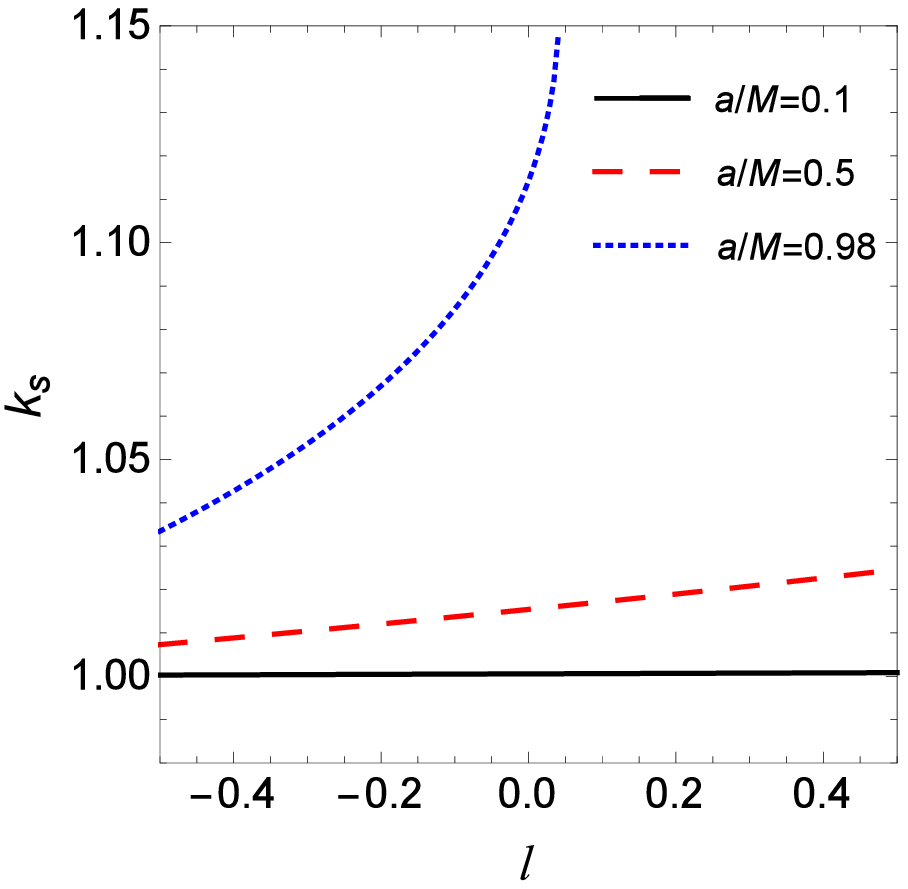}}
\end{center}
\caption{The ratio $k_{\rm s}$ of the shadow. (a)~$\theta=\frac{\pi}{6}$. (b)~$\theta=\frac{\pi}{4}$. (c)~$\theta=\frac{\pi}{2}$.}\label{ksfig}
\end{figure*}

For some different black holes, the observables $R_{\rm s}$ and $\delta_{\rm s}$ may have the same values, so we cannot distinguishes these black holes. Such phenomenon is known as the degeneracy. In order to eliminate it, more observables should be constructed. Here we would like to introduced two more distortion parameters $\epsilon$ and $k_{\rm s}$ \cite{NTsukamoto,RKumar,SWWei4}. To construct the first observable $\epsilon$, two new points $(x_{\rm h}, y_{\rm h})$ and ($x_{\rm c}, y_{\rm c}$) are needed. The former is just the point that the horizontal line of $y_{\rm h}=y_{\rm t}/2$ cuts the shadow at the side of $(x_{\rm l}, 0)$, and the latter is the center of the reference circle. The second observable denotes the ratio of the horizontal and vertical angular diameters of the shadow. These two observables can be calculated by the following formulas
\begin{eqnarray}
 \epsilon&=&1-\frac{\sqrt{(x_{\rm h}-x_{\rm c})^{2}+y_{\rm t}^{2}/4}}{R_{\rm s}},\\
 k_{\rm s}&=&\frac{2y_{\rm t}}{x_{\rm r}-x_{\rm l}}.
\end{eqnarray}
We display these two observables in Figs.~\ref{epsilonfig} and \ref{ksfig}, respectively. It is clear that both $\epsilon$ and $k_{\rm s}$ increase with $l$ for fixed $\theta$ and $a$. Moreover, we can also find that $\epsilon$ and $k_{\rm s}$ approach their maximum values for the extremal black holes, where the black hole spin takes its maximum value. This result implies that the shadow cast by the extremal black holes is most deformed.

\section{Photons in the presence of plasma}
\label{plasma}
Plasma is the most common phase of ordinary matter in the universe observed both in the accretion disk and the jet of the black hole. Hence it is important to investigate the influence of plasma on the motion of the photon. Such influence will also emerge in the shadow and deflection angle of the photon. Therefore, in this section, we aim to uncover these influences.

In Ref.~\cite{JLSynge2}, Synge gave a generalized expression of the refractive index $n$ of a static inhomogeneous plasma in the gravitational field
\begin{equation}
n(x^{i})^{2}=1+\frac{p_\mu p^\mu}{(p_\nu u^\nu)^2},\label{rindex}
\end{equation}
where $p_\mu$ and $u^\nu$ represent the four-momentum and four-velocity of the massless particle, respectively. Obviously, the vacuum case corresponds to $n(x^{i})=1$.

\begin{figure*}
\begin{center}
\includegraphics[width=8cm]{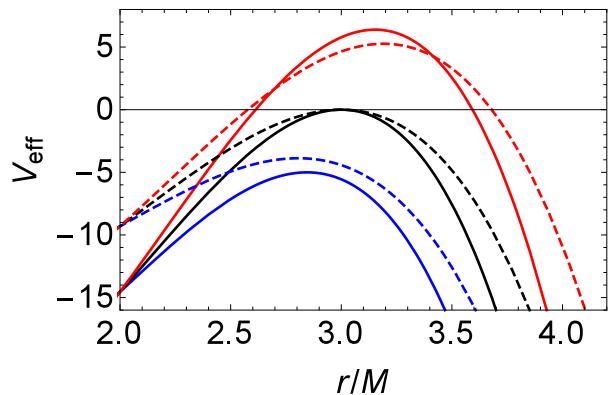}
\caption{The effective potential $V_{\rm eff}$ for the photon when $a=0$ and $l=0.1$. The red lines above, black lines in the middle and blue lines below indicate $\xi$=$\xi_c+0.2$, $\xi_c$, and $\xi_c-0.2$, respectively. The solid lines represent the vacuum $n$=1 and the dashed lines represent $n=0.8$. }\label{Vfig1}
\end{center}
\end{figure*}

\begin{figure*}
\begin{center}
\subfigure[]{\label{Veff1}
\includegraphics[width=8cm]{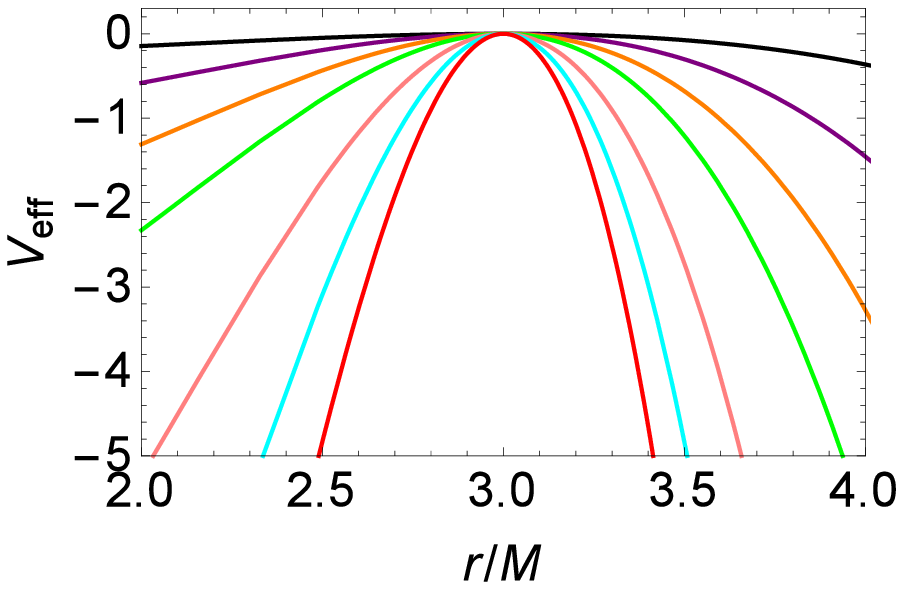}}~~~~~
\subfigure[]{\label{Veff2}
\includegraphics[width=8cm]{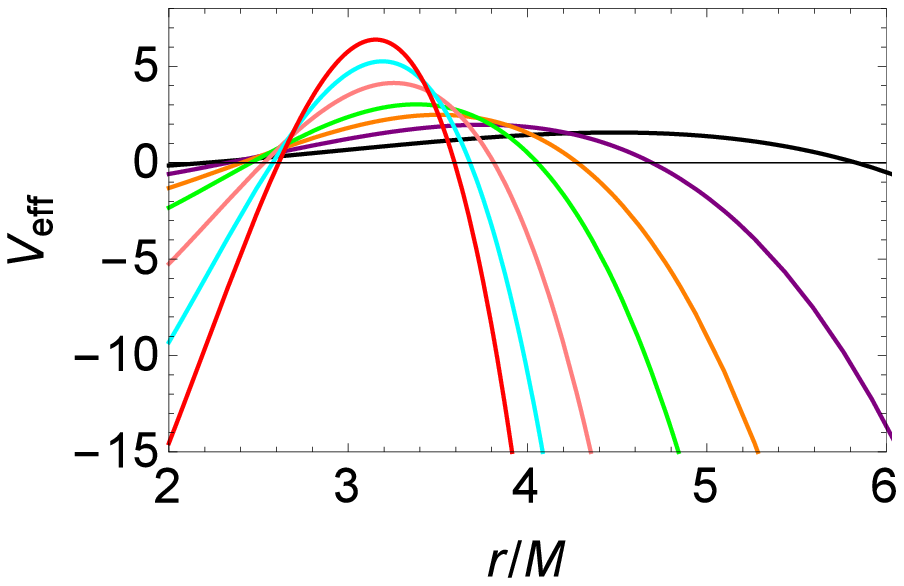}}
\caption{The effective potential $V_{\rm eff}$ as a function of $r$ when $a=0$ and $l=0.1$. From right to left (in the right side), the refractive index $n$ sets $0.1,~0.2,~0.3,~0.4,~0.6,~0.8,~1$, respectively. (a)~$\xi=\xi_c$. (b)~$\xi=\xi_c+0.2$.}\label{Vfig2}
\end{center}
\end{figure*}

Here we study the equation of motion of the photon orbiting the Kerr-like black hole with the presence of the plasma. The Hamiltonian for the photon is given by \begin{eqnarray}
H&=&\frac{1}{2}\left (g^{\mu\nu}p_{\mu}p_{\nu}
-(n(x^{i})^2-1)(p_t\sqrt{-g^{tt}})^2\right )\nonumber\\
&=&\frac{1}{2}\left (g^{tt}p_t^2n(x^{i})^2+g^{rr}p_r^2+g^{\vartheta\vartheta}p_\vartheta^2
+g^{\phi\phi}p_\phi^2+2g^{t\phi}p_tp_\phi\right ),\label{Hamiltonian}
\end{eqnarray}
and the solutions to the Hamilton equation are
\begin{equation}
\dot{p}_\mu = -\frac{\partial H}{\partial x^\mu},~~
\dot{x}^\mu = \frac{\partial H}{\partial p_\mu},\label{HamiltonianS}
\end{equation}
which determine the orbit of the photon. In this paper, we mainly consider the light rays in a static and homogeneous plasma, in which case the Hamilton-Jacobi equation is separable. In addition, the frequency of the photon is measured by an observer at infinity and could be considered as a constant. Therefore the refractive index becomes a constant, $n(x^{i})=\text{const.}$. Substituting \eqref{Hamiltonian} into \eqref{HamiltonianS}, we have
\begin{eqnarray}
(1+l)\rho^2\frac{dt}{d\lambda}\!&\!\!=\!\!&\!aL_{\rm z}(1+l)-a^2(1+l)^{\frac{3}{2}}E n^2 \sin^2\vartheta + \frac{r^2+(1+l)a^2}{\Delta} \left((r^2+(1+l)a^2)E n^2-aL_{\rm z}\sqrt{1+l} \right) ,\label{HS1}\\
 (1+l)\rho^2\frac{d\phi}{d\lambda}\!&\!\!=\!\!&\!\frac{(1+l)L_{\rm z}}{\sin^2\vartheta}
 -a(1+l)^{\frac{3}{2}}E+\frac{a\sqrt{1+l}}{\Delta} \left((r^2+(1+l)a^2)E-aL_{\rm z}\sqrt{1+l} \right)  ,\label{HS2}\\
 \rho^2\frac{dr}{d\lambda}\!&\!\!=\!\!&\!\sqrt{{\cal R}(r)},\label{HS3}\\
 \rho^2\frac{d\vartheta}{d\lambda}\!&\!\!=\!\!&\!\sqrt{\Theta(\vartheta)},\label{HS4}
 \end{eqnarray}
where
\begin{eqnarray}
{\cal R}(r)&=&-\Delta \left({\cal K}+(L_{\rm z}-aE\sqrt{1+l})^2 \right)
+ \left(\frac{\left(r^2+a^2(1+l)\right)En}{\sqrt{1+l}}-aL_{\rm z}\right)^2
+\frac{2aL_{\rm z}E\left(r^2+a^2(1+l)\right)}{\sqrt{1+l}}(n-1),\label{rrr}\\
\Theta(\vartheta)&=&{\cal K}+a^2E^2(1+l)(1-n^2\sin^2 \vartheta )
-L_{\rm z}^2 \cot^2 \vartheta.
\end{eqnarray}
In terms of the conserved parameters $\xi$ and $\eta$, the radial motion \eqref{HS3} can be reexpressed in the form of Eq.~\eqref{Veff} with the effective potential given by
 \begin{equation}
\frac{V_{\rm eff}}{E^2}=\left(\frac{r^2-2Mr}{1+l}+a^2 \right)
\left(\eta+(\xi-a\sqrt{1+l})^2\right)-\left(\frac{
 r^2+a^2(1+l)}{\sqrt{1+l}}-a\xi \right)^2-
 \frac{(n^2-1)\left(r^2+a^2(1+l)\right)^2}{1+l}.
\end{equation}
In Fig.~\ref{Vfig1}, we plot the effective potential $V_{\rm eff}$ for a non-rotating black hole with $l$=0.1. The refractive index $n$=1 and 0.8 for the solid and dashed curves. The conserved parameters $\xi$=$\xi_c+0.2$, $\xi_c$, and $\xi_c-0.2$ from top to bottom both for the solid and dashed curves. Since the allowed regions for the photons must have negative $V_{\rm eff}$, we observe that these photons coming from infinity with $\xi<\xi_c$ will be absorbed by the black hole. While these with $\xi>\xi_c$ will be reflected at the turning  points where $V_{\rm eff}$=0. A critical case occurs when $\xi=\xi_c$. On the other hand, we can also find that the refractive index $n$ influence the behavior of the effective potential $V_{\rm eff}$. However it has no influence on the radius of the photon sphere, where $V_{\rm eff}=\partial_r V_{\rm eff}$=0. To make it more clearly, the effective potential $V_{\rm eff}$ is plotted in Fig.~\ref{Vfig2} for $n$=0.1, 0.2, 0.3, 0.4, 0.6, 0.8, and 1. When $\xi=\xi_c$, it is easy to find that the radius of the photon sphere is always at $3M$ independent of $n$. While for $\xi>\xi_c$, the turning point is shifted to a large distance for small $n$.

\subsection{Shadow}
\label{shadowplasma}

The boundary of the shadow is constructed by the circular null orbit determined by
\begin{eqnarray}
{\cal R}(r)=0,\quad \frac{d{\cal R}(r)}{dr}=0, \quad \text{and} \quad \frac{d^2{\cal R}(r)}{dr^2}<0.
\end{eqnarray}
Employing \eqref{rrr}, we find that for a non-rotating black hole, the radius of the photon sphere $r_{\text{ps}}=3M$ is independent of $l$ and $n$. Further, these two conserved parameters $\xi$ and $\eta$ describing the profile of the shadow satisfy
 \begin{equation}
\eta +\xi^2=\frac{n^2r_{\text{ps}}^3}{-2M+r_{\text{ps}}}.
\label{Sch}
\end{equation}
Insert $r_{\text{ps}}=3M$ into above equation, the radius of the shadow reads
\begin{equation}
 R_{\rm s}=3n\sqrt{3}M. \label{Schradius}
 \end{equation}
So we can conclude that the plasma has no effect on the photon sphere but on the radius of the shadow. In particular, $R_{\rm s}$ is proportional to the refractive index $n$.

\begin{figure*}
\begin{center}
\subfigure[]{\label{shadow1}
\includegraphics[width=5cm]{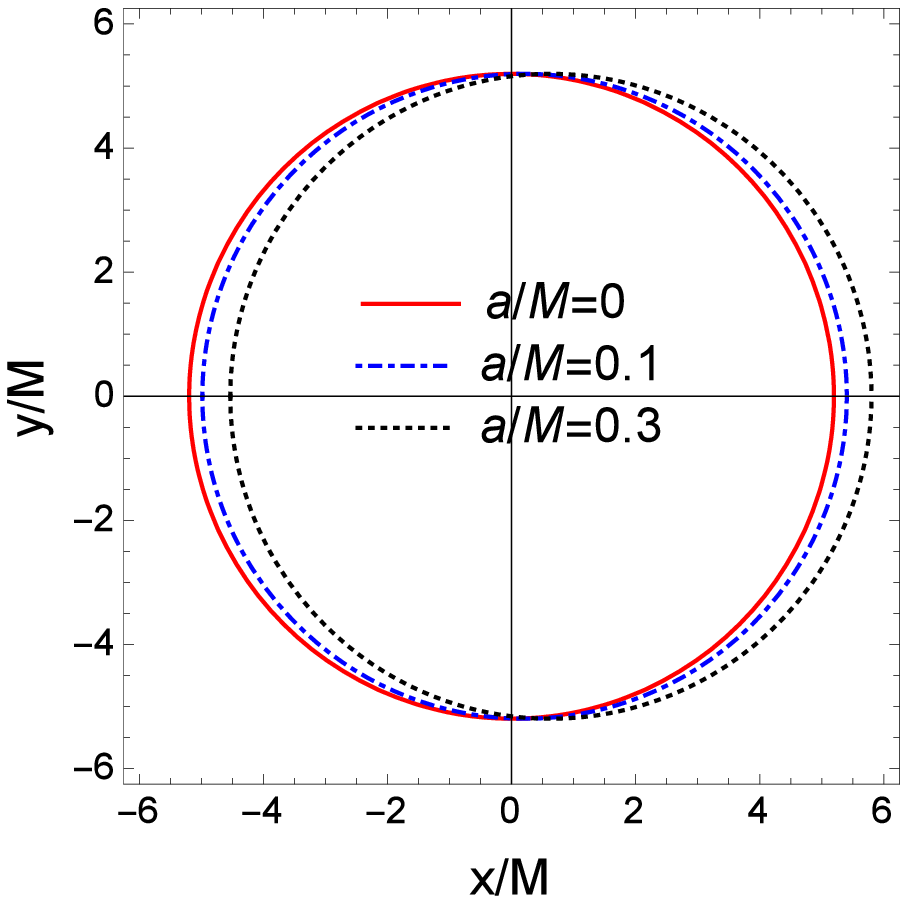}}~~~~~
\subfigure[]{\label{shadow2}
\includegraphics[width=5cm]{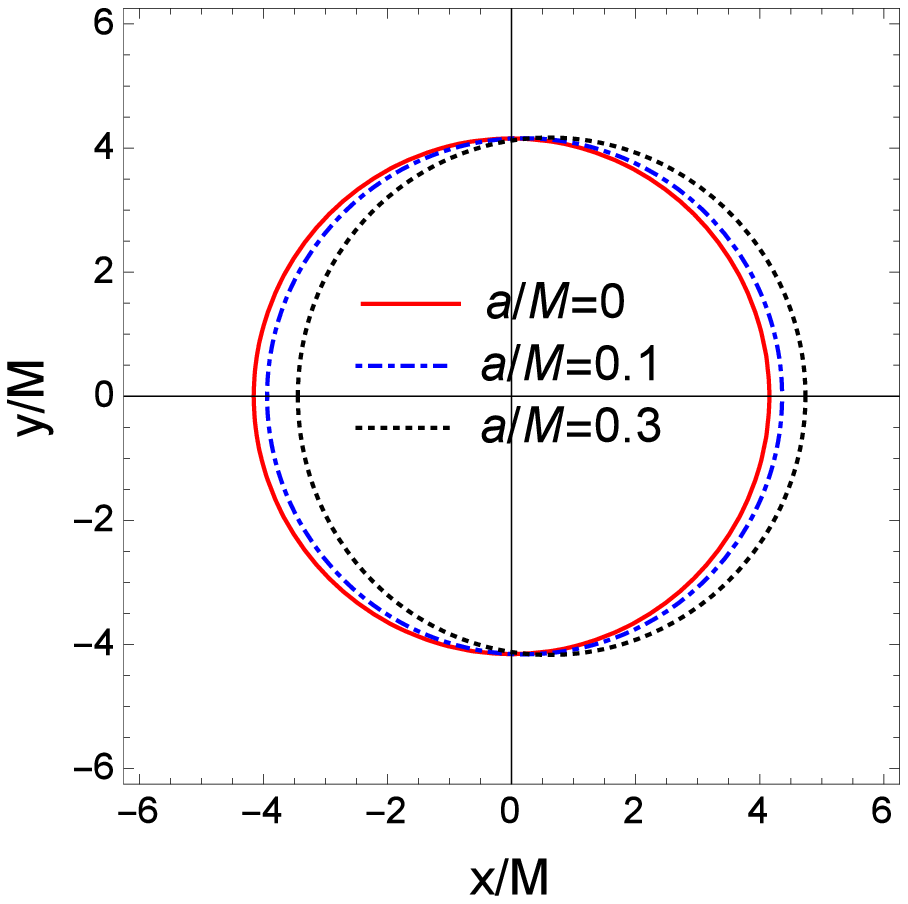}}~~~~~
\subfigure[]{\label{shadow3}
\includegraphics[width=5cm]{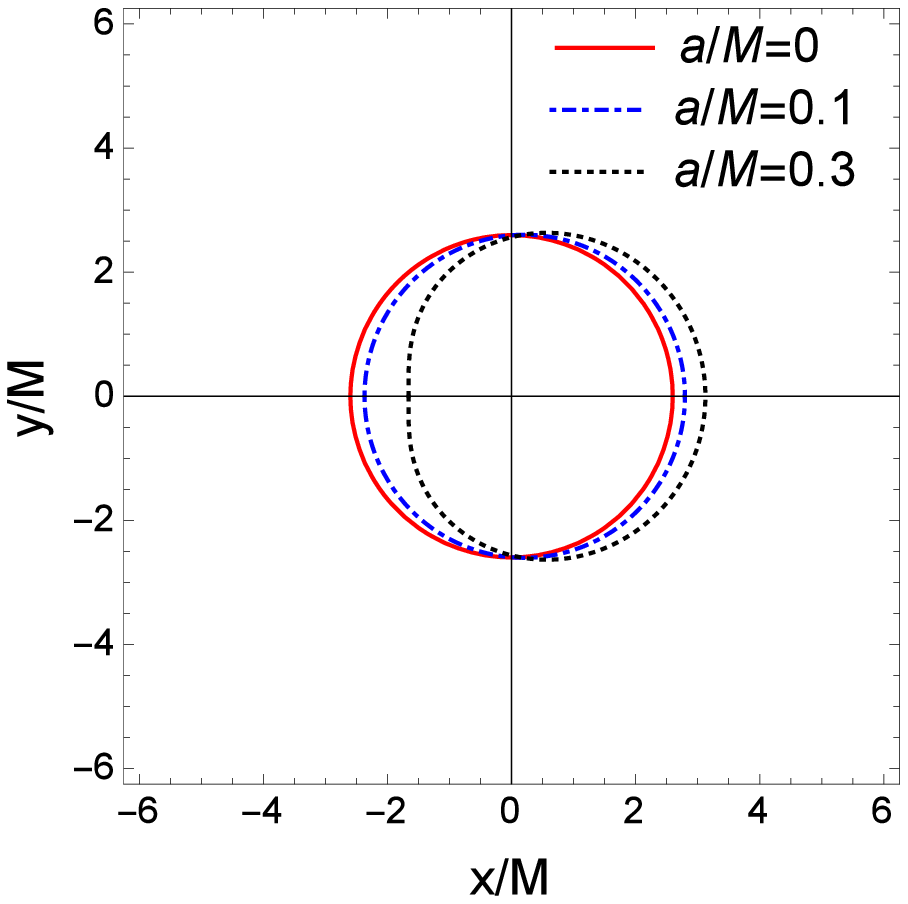}}
\caption{Shadows for the Kerr-like black hole with plasma when  $\theta=\pi/2$, $l=0.1$. (a)~$n=1$. (b)~$n=0.8$. (c)~$n=0.5$.}\label{shadowfig}
\end{center}
\end{figure*}

For the rotating black hole case, one has
\begin{eqnarray}
 \xi&=&\frac{M\left( a^2(1+l)-r^2 \right)+\sqrt{X(r)}}{a\sqrt{1+l}(M-r)},\\
 \eta&=&\frac{a^4(1\!+\!l)^2(n^2\!-\!1)(M\!-\!r)^2
 -\!2a^2(1\!+\!l)Mr^2(M(n^2\!-\!1)-\!n^2r)-\!r^4(M^2(2\!+\!3n^2)
 -\!4Mn^2r+\!n^2r^2)+\!2r^2\!\sqrt{X(r)}}{a^2(1+l)(M-r)^2},
\end{eqnarray}
where
\begin{eqnarray}
X(r)&=&-a^4(l+1)^2 \left(M^2 \left(n^2-1\right)-n^2 r^2\right)+2 a^2 (l+1) r^2 (M (n-1)-n r) (M n+M-n r) \\ \nonumber
&~&+r^4 \left(3 M^2 n^2+M^2-4 M n^2 r+n^2 r^2\right).
\end{eqnarray}
Here we will restrict our attention to the case that the observer is on the equatorial plane, i.e., $\theta=\pi/2$. Then one can easily get
\begin{eqnarray}
 x=-\xi,\quad
 y=\pm\sqrt{\eta}.
\end{eqnarray}
The black hole shadow shapes are described in Fig.~\ref{shadowfig} with a small Lorentz-violation $l=0.1$. For $n$=1, 0.8, and 0.5, the results are, respectively, shown in Figs.~\ref{shadow1}-\ref{shadow3}. Obviously, for $a$=0, the shadows are standard circles, and their size shirks for small $n$. On the other hand, for fixed $n$, each shape is shifted toward to the right by the black hole spin $a$. A significant behavior is that, when $n$=0.5, the shadow gets an obvious deformed, see the dashed contour given in Fig.~\ref{shadow3}. So the plasma can reinforce the deformation of the shadow.

\subsection{Deflection angle}
\label{deflection}

When $\xi>\xi_c$, the photon from infinity will be bounced back to infinity at a certain distance. For a distant observer, the photon will not go along a straight line, but be curved by the spacetime. In this section, we would like to consider the deflection angle of a photon passing by the black hole.

First of all, let us reconsider the refractive index of the plasma. For a cold and non-magnetized plasma, the frequency $\omega_p$ is defined as
 \begin{equation}
\omega_p(r)^2=\frac{4\pi e^2}{m}N(r),
\end{equation}
where $e$, $m$, and $N(r)$ are the charge, mass, and the number density of the electrons, respectively. Again, for a homogeneous plasma, $N(r)$ is independent of the radius $r$, and naturally $\omega_p$ is a constant. Here we should remind that $\omega$ is the frequency of the photon measured by a static observer at infinity. It is worth noting that if $\omega < \omega_p$, the light will not be able to propagate and be evanescent, so only when $\omega > \omega_p$, the light can spread smoothly in the plasma environment. Then the refractive index of the plasma can be rewritten as
  \begin{equation}
n^2=1-\frac{\omega_p^2}{\omega^2}.
\end{equation}
The Hamiltonian \eqref{Hamiltonian} for the photon in the plasma also reads
\begin{eqnarray}
H=\frac{1}{2}(g^{\mu\nu}p_{\mu}p_{\nu}+\omega_p^2). \label{Hamiltonian2}
\end{eqnarray}
For simplicity, we set $\theta=\pi/2$. According to Eqs.~\eqref{HamiltonianS} and \eqref{Hamiltonian2}, we know $\dot{p}_t=\dot{p}_\phi=0$, which means $p_t$ and $p_\phi$ are constants for the motion. For an asymptotically flat spacetime, we have $g_{tt}\rightarrow -1$ when $r\rightarrow \infty$. By the gravitational redshift, $\omega$ will be the form of
\begin{eqnarray}
\omega(\infty)=\frac{\omega_0}{\sqrt{-g_{tt}(\infty)}}=\omega_0,
\end{eqnarray}
where $\omega_0\equiv -p_t$. Besides, from the motion equations, we obtain
\begin{eqnarray}
\frac{dr}{d\phi}=\frac{\dot{r}}{\dot{\phi}}=\frac{g^{rr} p_r }{g^{\phi \phi}p_\phi+g^{t \phi}p_t},
\end{eqnarray}
and taking advantage of $H=0$, we can rewrite it as
\begin{eqnarray}
\frac{dr}{d\phi}=\sqrt{\frac{g^{rr}}{g^{\phi \phi}}}\sqrt{\frac{\omega_0^2}{(p_\phi-\frac{g^{t\phi}}{g^{\phi \phi}}\omega_0)^2} h(r)^2-1},\label{dr}
\end{eqnarray}
where $h(r)^2$ is given by
\begin{equation}
h(r)^2 = -\frac{g^{tt}}{g^{\phi \phi}}+\frac{g^{t\phi}g^{t\phi}}{{g^{\phi \phi} g^{\phi \phi}}}-\frac{\omega_p^2}{g^{\phi \phi}\omega_0^2}.
\end{equation}
As mentioned above, the photon from infinity will be deflected near the black hole, and then escapes to infinity when $\xi>\xi_c$. At the minimum distance $R$, $dr/d\phi=0$ satisfies, and therefore we have
\begin{equation}
h(R)^2 = \frac{(p_\phi-\frac{g^{t\phi}}{g^{\phi \phi}}\omega_0)^2}{\omega_0^2}.
\end{equation}
Integrating Eq.~\eqref{dr} with respect to $r$, we get the representation of the deflection angle of light ray in the Kerr-like spacetime
\begin{eqnarray}
\hat{\alpha}&=&2\int_R^\infty \sqrt{\frac{g^{\phi \phi}}{g^{rr}}}\left( \frac{h(r)^2}{h(R)^2}-1\right)^{-1/2}dr-\pi\nonumber\\
&=&2\int_R^\infty \frac{\sqrt{-(1+l)(2M-r)r}}{a^2(1+l)+r(-2M+r)}
\bigg(\frac{\frac{r\left(a^2(1+l)+r(-2M+r)\right) }{2M-r}\left(\frac{r}{2M-r}+\frac{\omega_p^2}{\omega_0^2} \right)}{\frac{R\left(a^2(1+l)+R(-2M+R)\right) }{2M-R}\left(\frac{R}{2M-R}+\frac{\omega_p^2}{\omega_0^2} \right)}-1
\bigg)^{-1/2}dr-\pi,\label{defelctionangleK}
\end{eqnarray}
where the frequency $\omega_p$ vanishes when the plasma is absent. Especially, for a non-rotating black hole, the above expression~\eqref{defelctionangleK} reduces to
\begin{eqnarray}
\hat{\alpha}=2\int_R^\infty \sqrt{\frac{1+l}{r(-2M+r)}}
\bigg(\frac{\frac{r^3}{-2M+r}-\frac{r^2\omega_p^2}{\omega_0^2}}
{\frac{R^3}{-2M+R}-\frac{R^2\omega_p^2}{\omega_0^2}}-1
\bigg)^{-1/2}dr-\pi.\label{defelctionangleS}
\end{eqnarray}
For given a series of the parameters, one can numerically obtain the deflection angle of a photon.

\section{Observational constraints}
 \label{constraint}

\begin{figure*}
\begin{center}
\includegraphics[width=5cm]{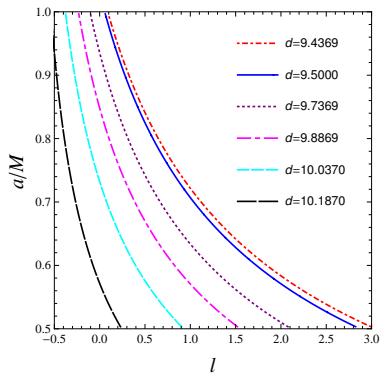}
\caption{Contours of the diameter $d$ in $l-a/M$ plane.} \label{diameterfig}
\end{center}
\end{figure*}

Based on the bending of light and the capture of photons at the event horizon, EHT reconstructs the first event-horizon-scale image of the supermassive black hole candidate in the center of the giant elliptical galaxy M87, and gives a central mass $M=(6.5\pm0.7\mid_{sys}\pm0.2\mid_{stat})\times10^9M_\odot$ ($M_\odot$ is the mass of the Sun) by comparing images to an extensive library of ray-traced general-relativistic magnetohydrodynamic simulations of black holes. The distance to earth is also shown as $D=16.8^{+0.8}_{-0.7}Mpc$. Besides, they measure the angular diameter of $\theta_{\rm s}=(42\pm3)\mu as$ \cite{KAkiyama1,KAkiyama2,KAkiyama3,KAkiyama4,KAkiyama5,KAkiyama6}. And these observations indicate that the range of the spin parameter of M87$^*$ is $(0.5,0.94)$. Meanwhile, depending on the angle of the jet, the rotation axis should be about $17^{\circ}$ from the observer \cite{RCraigWalker}. In the following discussion, we will constrain the parameters in this paper by theses observational data. Here we adopt the size of M87$^*$ in unit mass, which is given in Ref.~\cite{CBambi}
\begin{equation}
d_{M87^*}\equiv\frac{D\theta_{\rm s}}{M}\simeq 11.0\pm1.5.\label{m887}
\end{equation}
Considering the observation angle $\theta=17^{\circ}$ and the range of $a$, we calculate the corresponding diameter $d$ with the spin parameter $a$ and the LV parameter $l$. The numerical results are exhibited in Fig.~\ref{diameterfig}. For each contour of the diameter $d$, $a$ decreases with $l$. Considering the error, the blue solid line in the figure represents the minimum value that the diameter can take. In other words, the region corresponding to the left of the blue line is allowed, which tells that the possible ranges of the $a$ and $l$ are still very large. Moreover, Table~\ref{diametertable} shows the maximum values of the LV parameter $l$ that can be obtained as $a$ fixed. For instance, the allowed range of $l$ is $(-0.5, 2.873)$ when $a=0.5$, while the range shrinks to $(-0.5, 0.132)$ when $a$ takes the maximum value 0.94.

\begin{table*}
\begin{tabular}{p{1cm}|p{0.5cm}p{1.5cm}|p{0.5cm}p{1.5cm}|p{0.5cm}p{1.5cm}|p{0.5cm}p{1.5cm}|p{0.5cm}p{1.5cm}|p{0.5cm}p{1.2cm}}
\hline
~~$a$ & &0.5 & &0.6  & &0.7   & &0.8&   &0.9&   &0.94\\ \hline
~~$l $  &  & 2.873 &  &1.728 & & 1.037& &0.563& &0.235& &0.132\\
\hline
\end{tabular}
\caption{The maximum values that can be obtained by the LV parameter $l$ with $a$ fixed.}\label{diametertable}
\end{table*}

On the other hand, Eq.~\eqref{Schradius} shows that when a plasma is present, the shadow radius $R_{\rm s}$ is independent of the LV parameter $l$, but dependent of the refractive index $n$ for a non-rotating black hole. Since the black hole spin has less influence on the shadow size, we can constrain the refractive index $n$. Considering the diameter \eqref{m887}, we find $n$ is favored in the following region
\begin{equation}
0.914 \leqslant n \leqslant 1.
\end{equation}

\section{Conclusion}
 \label{Conclusion}

We have studied the detailed optical properties of the shadow cast by a Kerr-like black hole in the Einstein-bumblebee gravity with and without plasma.

At first, in the background of the Kerr-like black hole, we obtained the null geodesics via the Hamilton-Jacobi equation. By making the use of the unstable circular null orbit conditions, two conserved parameters $\xi$ and $\eta$ determining the motion of the photon were obtained. Based on them, the shadow is obtained  with the celestial coordinates. Then we studied the size and deformation of the shadow. When taking the observation angle $\theta=\pi/6,\pi/4$, and $\pi/2$, we calculated the radius $R_{\rm s}$ of the shadow. for different value parameter space. For fixed spin $a$ and LV parameter $l$, we found that the observers at lower latitudes will always see a larger shadow. When $\theta=\pi/6$ and $\pi/4$, a slowly rotating black hole has a larger shadow than a fast one. And $R_{\rm s}$ always decreases with $l$. However, when $\theta$ near $\pi/2$, $R_{\rm s}$ increases with $l$. In particular, the rapidly rotating black hole corresponds to a larger shadow for the same $l$. Then we investigated two distortion parameters $\delta_{\rm s}$ and $\epsilon$, as well as the ratio parameter $k_{\rm s}$ for the shadow. The results implied that all these three observables increase with $l$.

Moreover, we examined the influence of plasma on the motion of the photon. In a homogeneous plasma, the geodesics of the photon was obtained following the Hamilton-Jacobi approach. Considering three cases $\xi<\xi_c$, $\xi=\xi_c$, and $\xi>\xi_c$, we figured out that the plasma has no influence on the radius of the photon sphere, but it shifts the turning point of the photon. In addition, the shadow shape is exhibited in Fig.~\ref{shadowfig}. The results showed that the size of shadow increases with the refractive index $n$ and the plasma reinforces the deformation of the shadow.

Finally, employing the observation of M87*, we constrained the LV parameter $l$. Via the angular diameter of shadow, we figured out the allowed region in the parameter space. Furthermore, after considering the influence of the refractive index $n$, we uncovered that the region $n\in$(0.914, 1) is more favored.

\section*{Acknowledgements}
We thank Dr. Zi-Chao Lin for helpful discussions. This work was supported by the National Natural Science Foundation of China (Grants No. 12075103 and No. 11675064), the 111 Project (Grant No. B20063), and the Fundamental Research Funds for the Central Universities (No. Lzujbky-2019-ct06).


\begin{thebibliography}{99}


 \bibitem{KAkiyama1}
     K.~Akiyama \emph{et al.} [Event Horizon Telescope Collaboration],
  {\em First M87 Event Horizon Telescope results. I. The shadow of the supermassive black hole},
   Astrophys. J. \textbf{875}, L1 (2019) [arXiv:1906.11238 [astro-ph.GA]].

\bibitem{KAkiyama2}
     K.~Akiyama \emph{et al.} [Event Horizon Telescope Collaboration],
  {\em First M87 Event Horizon Telescope results. II. Array and instrumentation},
   Astrophys. J. \textbf{875}, L2 (2019) [arXiv:1906.11239 [astro-ph.IM]].

\bibitem{KAkiyama3}
     K.~Akiyama \emph{et al.} [Event Horizon Telescope Collaboration],
  {\em First M87 Event Horizon Telescope results. III. Data processing and calibration},
   Astrophys. J. \textbf{875}, L3 (2019) [arXiv:1906.11240 [astro-ph.GA]].

\bibitem{KAkiyama4}
     K.~Akiyama \emph{et al.} [Event Horizon Telescope Collaboration],
  {\em First M87 Event Horizon Telescope results. IV. Imaging the central supermassive black hole},
   Astrophys. J. \textbf{875}, L4 (2019) [arXiv:1906.11241 [astro-ph.GA]].

\bibitem{KAkiyama5}
     K.~Akiyama \emph{et al.} [Event Horizon Telescope Collaboration],
  {\em First M87 Event Horizon Telescope results. V. Physical origin of the asymmetric ring},
   Astrophys. J. \textbf{875}, L5 (2019) [arXiv:1906.11242 [astro-ph.GA]].

\bibitem{KAkiyama6}
     K.~Akiyama \emph{et al.} [Event Horizon Telescope Collaboration],
  {\em First M87 Event Horizon Telescope results. VI. The shadow and mass of the central black hole},
   Astrophys. J. \textbf{875}, L6 (2019) [arXiv:1906.11243 [astro-ph.GA]].

\bibitem{JLSynge1}
 J.L.~Synge,
 {\em The escape of photons from gravitationally intense stars},
   Mon. Not. Roy. Astron. Soc. \textbf{131}, 463-466 (1966).

\bibitem{JPLuminet}
 J.P.~Luminet,
 {\em Image of a spherical black hole with thin accretion disk},
   A$\&$A \textbf{75}, 228 (1979).

\bibitem{JMBardeen}
 J.M.~Bardeen,
 {\em Timelike and null geodesics of the Kerr metric},
  Gordon Breach, Science Publishers, New York (1973).


  \bibitem{CBambi1}
 C.~Bambi and K.~Freese,
   {\em Apparent shape of super-spinning black holes},
     Phys. Rev. D \textbf{79}, 043002 (2009) [arXiv:0812.1328 [astro-ph]].


 \bibitem{KHioki}
 K.~Hioki and K.~Maeda,
  {\em Measurement of the Kerr spin parameter by observation of a compact object's shadow},
  Phys. Rev. D \textbf{80}, 024042 (2009) [arXiv:0904.3575 [astro-ph.HE]].

 \bibitem{CBambi2}
    C.~Bambi and N.~Yoshida,
   {\em Shape and position of the shadow in the $\delta$=2 Tomimatsu-Sato space-time},
      Class. Quant. Grav. \textbf{27}, 205006 (2010) [arXiv:1004.3149 [gr-qc]].

\bibitem{ZLLi}
    Z.L.~Li and C.~Bambi,
      {\em Measuring the Kerr spin parameter of regular black holes from their shadow},
      JCAP \textbf{1401}, 041 (2014) [arXiv:1309.1606 [gr-qc]].


  \bibitem{NTsukamoto}
 N.~Tsukamoto, Z.~Li and C.~Bambi,
  {\em Constraining the spin and the deformation parameters from the black hole shadow}, JCAP \textbf{1406}, 043 (2014) [arXiv:1403.0371 [gr-qc]].

\bibitem{TJohannsen}
 T.~Johannsen,
  {\em Photon rings around Kerr and Kerr-like black holes},
  Astrophys. J. \textbf{777}, 170 (2013) [arXiv:1501.02814 [astro-ph.HE]].

\bibitem{AAbdujabbarov1}
 A.~Abdujabbarov, L.~Rezzolla, and B.~Ahmedov,
  {\em A coordinate-independent characterization of a black hole shadow},
   Mon. Not. Roy. Astron. Soc. \textbf{454}, 2423 (2015) [arXiv:1503.09054 [gr-qc]].

\bibitem{MGhasemi}
 M.~Ghasemi-Nodehi, Z.L.~Li, and C.~Bambi,
  {\em Shadows of CPR black holes and tests of the Kerr metric},
   Eur. Phys. J. C \textbf{75}, 315 (2015) [arXiv:1506.02627 [gr-qc]].

\bibitem{RKumar}
 R.~Kumar and S.~G.~Ghosh,
  {\em Black Hole Parameter Estimation from Its Shadow},
   Astrophys. J. \textbf{892}, 78 (2020) [arXiv:1811.01260 [gr-qc]].


  \bibitem{SWWei1}
 S.W.~Wei, P.~Cheng, Y.~Zhong, and X.N.~Zhou,
  {\em Shadow of noncommutative geometry inspired black hole},
   JCAP \textbf{1508}, 004 (2015) [arXiv:1501.06298 [gr-qc]].

\bibitem{SAbdolrahimi}
 S.~Abdolrahimi, R.B.~Mann, and C.~Tzounis,
  {\em Distorted local shadows},
  Phys. Rev. D \textbf{91}, 084052 (2015) [arXiv:1502.00073 [gr-qc]].


\bibitem{Aovgun1}
  A.~\"{o}vg\"{u}n, \.{I}.~Sakall{\i} and J.~Saavedra,
  {\em Shadow cast and deflection angle of Kerr-Newman-Kasuya spacetime},
   JCAP \textbf{1810}, 041 (2018) [arXiv:1807.00388 [gr-qc]].

\bibitem{HMWang}
  H.M.~Wang, Y.M.~Xu and S.W.~Wei,
  {\em Shadows of Kerr-like black holes in a modified gravity theory},
   JCAP \textbf{1903}, 046 (2019) [arXiv:1810.12767 [gr-qc]].

\bibitem{SWWei2}
 S.W.~Wei, Y.X.~Liu, and R.B.~Mann,
  {\em Intrinsic curvature and topology of shadows in Kerr spacetime},
  Phys. Rev. D \textbf{99}, 041303 (2019) [arXiv:1811.00047 [gr-qc]].


\bibitem{SWWei3}
 S.W.~Wei, Y.C.~Zou, Y.X.~Liu, and R.B.~Mann,
  {\em Curvature radius and Kerr black hole shadow},
  JCAP \textbf{1908}, 030 (2019) [arXiv:1904.07710 [gr-qc]].

\bibitem{MZWang}
 M.Z.~Wang, S.B.~Chen, J.C.~Wang and J.L.~Jing,
  {\em Shadow of a Schwarzschild black hole surrounded by a Bach-Weyl ring},
  Eur. Phys. J. C \textbf{80}, 110 (2020) [arXiv:1904.12423 [gr-qc]].

\bibitem{TZhu}
  T.~Zhu, Q.~Wu, M.~Jamil and K.~Jusufi,
  {\em Shadows and deflection angle of charged and slowly rotating black holes in Einstein-{\AE}ther theory},
  Phys. Rev. D \textbf{100}, 044055 (2019) [arXiv:1906.05673 [gr-qc]].


\bibitem{CLiu1}
  C.~Liu, C.~Ding and J.~Jing,
  {\em Thin accretion disk around a rotating Kerr-like black hole in Einstein-bumblebee gravity model}, (2019), arXiv:1910.13259 [gr-qc].

\bibitem{CLiu2}
 C.~Liu, T.~Zhu, Q.~Wu, K.~Jusufi, M.~Jamil, M.~Azreg-A\"{\i}nou and A.~Wang,
  {\em Shadow and quasinormal modes of a rotating loop quantum black hole}, Phys. Rev. D \textbf{101}, 084001 (2020) [arXiv:2003.00477 [gr-qc]].

\bibitem{SWWei4}
 S.W.~Wei and Y.X.~Liu,
  {\em Testing the nature of Gauss-Bonnet gravity by four-dimensional rotating black hole shadow}, Eur. Phys. J. Plus \textbf{136}, 436 (2021) [arXiv:2003.07769 [gr-qc]].

\bibitem{MKhodadi}
 M.~Khodadi and E.~N.~Saridakis,
  {\em Einstein-\AE ther Gravity in the light of Event Horizon Telescope Observations of M87*}, Phys. Dark Univ.  \textbf{32}, 100835 (2021) [arXiv:2012.05186 [gr-qc]].

\bibitem{Ichimaru}
  S.~Ichimaru,
  {\em Bimodal behavior of accretion disks - Theory and application to Cygnus X-1 transitions}, Astrophys. J. \textbf{214}, 840 (1977).

\bibitem{Narayan}
  R.~Narayan and I.~Yi,
  {\em Advection dominated accretion: Underfed black holes and neutron stars},
  Astrophys. J. \textbf{452}, 710 (1995) [astro-ph/9411059].

\bibitem{Hollywood}
  J.M.~Hollywood and F.~Melia,
  {\em General relativistic effects on the infrared spectrum of thin accretion disks in active galactic nuclei : Application to Sagittarius A*}, Astrophys. J. Suppl. \textbf{112}, 423 (1997).
















\bibitem{RCThomson}
  R.C.~Thomson, D.R.T.~Robinson, N.R.~Tanvir, C.D.~Mackay and A.~Boksenberg,
  {\em Hst polarization map of the ultraviolet emission from the outer jet in m87 and a comparison with the 2cm radio emission}, Mon. Not. Roy. Astron. Soc. \textbf{275}, 921 (1995) [astro-ph/9505121].

\bibitem{CSReynolds}
  C.S.~Reynolds, A.C.~Fabian, A.~Celotti and M.~J.~Rees,
  {\em The matter content of the jet in m87: evidence for an electron-positron jet},
  Mon. Not. Roy. Astron. Soc. \textbf{283}, 873 (1996) [astro-ph/9603140].

\bibitem{YYKovalev}
  Y.Y.~Kovalev, M.L.~Lister, D.C.~Homan and K.I.~Kellermann,
  {\em The Inner Jet of the Radio Galaxy M87},
  Astrophys. J. Lett. \textbf{668}, L27 (2007)
  [arXiv:0708.2695 [astro-ph]].




 \bibitem{ABroderick}
   A.~Broderick and R.~Blandford,
  {\em Covariant magnetoionic theory - I. Ray propagation},
   Mon. Not. Roy. Astron. Soc. \textbf{342}, 1280 (2003) [astro-ph/0302190].

\bibitem{VPerlick1}
   V.~Perlick, O.Y.~Tsupko and G.S.~Bisnovatyi-Kogan,
  {\em Influence of a plasma on the shadow of a spherically symmetric black hole},
   Phys. Rev. D \textbf{92}, 104031 (2015) [arXiv:1507.04217 [gr-qc]].

\bibitem{FAtamurotov}
  F.~Atamurotov and B.~Ahmedov,
  {\em Optical properties of black hole in the presence of plasma: shadow},
   Phys. Rev. D \textbf{92}, 084005 (2015) [arXiv:1507.08131 [gr-qc]].


\bibitem{AAbdujabbarov2}
   A.~Abdujabbarov, B.~Toshmatov, Z.~Stuchl\'{\i}k and B.~Ahmedov,
  {\em Shadow of the rotating black hole with quintessential energy in the presence of plasma},
   Int.\ J.\ Mod.\ Phys.\ D \textbf{26}, 1750051 (2016) [arXiv:1512.05206 [gr-qc]].

   \bibitem{CQLiu}
   C.Q.~Liu, C.K.~Ding and J.L.~Jing,
  {\em Effects of homogeneous plasma on strong gravitational lensing of Kerr black holes},
   Chin.\ Phys.\ Lett. \textbf{34}, 090401 (2017) [arXiv:1610.02128 [gr-qc]].


   \bibitem{VPerlick2}
   V.~Perlick, O.Y.~Tsupko,
  {\em Light propagation in a plasma on Kerr spacetime: Separation of the Hamilton-Jacobi equation and calculation of the shadow},
   Phys. Rev. D \textbf{95}, 104003 (2017) [arXiv:1702.08768 [gr-qc]].

    \bibitem{HYan}
   H.~Yan,
  {\em Influence of a plasma on the observational signature of a high-spin Kerr black hole},
   Phys. Rev. D \textbf{99}, 084050 (2019) [arXiv:1903.04382 [gr-qc]].

\bibitem{SDastan}
   S.~Dastan, R.~Saffari and S.~Soroushfar,
  {\em Shadow of a charged rotating black hole in $f(R)$ gravity}, (2016), arXiv:1606.06994 [gr-qc].

 \bibitem{ASaha}
    A.~Saha, S.M.~Modumudi and S.~Gangopadhyay,
  {\em Shadow of a noncommutative geometry inspired Ay\'{o}n Beato Garc\'{\i}a black hole},
   Gen. Rel. Grav. \textbf{50}, 103 (2018) [arXiv:1802.03276 [gr-qc]].

\bibitem{ADas}
  A.~Das, A.~Saha and S.~Gangopadhyay,
  {\em Shadow of charged black holes in Gauss-Bonnet gravity},
   Eur. Phys. J. C \textbf{80}, 180 (2020) [arXiv:1909.01988 [gr-qc]].


\bibitem{GZBabar}
  G.Z.~Babar, A.Z.~Babar and F.~Atamurotov,
  {\em Optical properties of Kerr-Newman spacetime in the presence of plasma},
   Eur. Phys. J. C \textbf{80}, 761 (2020) [arXiv:2008.05845 [gr-qc]].

   \bibitem{MFathi}
  M.~Fathi and J.R.~Villanueva,
  {\em The role of elliptic integrals in calculating the gravitational lensing of a charged Weyl black hole surrounded by plasma}, (2020) [arXiv:2009.03402 [gr-qc]].

\bibitem{Mattingly}
  D. Mattingly,
  {\em Modern tests of Lorentz invariance}, Living Rev. Rel. \textbf{8}, 5 (2005) [arXiv:gr-qc/0502097].


\bibitem{CRovelli}
 C.~Rovelli,
  {\em Ashtekar formulation of general relativity and loop space nonperturbative quantum gravity: A Report},
   Class. Quant. Grav. \textbf{8}, 1613 (1991).

\bibitem{AAshtekar2}
 A.~Ashtekar, C.~Rovelli and L.~Smolin,
  {\em Gravitons and loops},
  Phys. Rev. D \textbf{44}, 1740 (1991) [hep-th/9202054].

\bibitem{AAshtekar1}
 A.~Ashtekar and C.~Rovelli,
  {\em A Loop representation for the quantum Maxwell field},
   Class. Quant. Grav. \textbf{9}, 1121 (1992) [hep-th/9202063].


\bibitem{AAshtekar3}
 A.~Ashtekar, C.~Rovelli and L.~Smolin,
  {\em Weaving a classical geometry with quantum threads},
   Phys. Rev. Lett. \textbf{69}, 237 (1992) [hep-th/9203079].

\bibitem{AAshtekar4}
 A.~Ashtekar,
  {\em Recent developments in classical and quantum theories of connections including general relativity}, (1992) hep-th/9205038.

\bibitem{VAKostelecky1}
  V.A.~Kosteleck\'{y} and S.~Samuel,
  {\em Spontaneous breaking of Lorentz symmetry in string theory}, Phys. Rev. D \textbf{39}, 683 (1989).

\bibitem{VAKostelecky2}
  V.A.~Kosteleck\'{y} and S.~Samuel,
  {\em Gravitational phenomenology in higher dimensional theories and strings}, Phys. Rev. D \textbf{40}, 1886 (1989).

  \bibitem{VAKostelecky3}
  V.A.~Kosteleck\'{y} and S.~Samuel,
  {\em Phenomenological Gravitational constraints on strings and higher dimensional theories}, Phys. Rev. Lett. \textbf{63}, 224 (1989).

  \bibitem{VAKostelecky4}
  V.A.~Kosteleck\'{y} and R.~Lehnert,
  {\em Stability, causality, and Lorentz and CPT violation}, Phys. Rev. D \textbf{63}, 065008 (2001) [hep-th/0012060].

  \bibitem{VAKostelecky5}
  V.A.~Kosteleck\'{y},
  {\em Gravity, Lorentz violation, and the standard model}, Phys. Rev. D \textbf{69}, 105009 (2004) [hep-th/0312310].

  \bibitem{RBluhm1}
  R.~Bluhm and V.A.~Kostelecky,
  {\em Spontaneous Lorentz violation, Nambu-Goldstone modes, and gravity},
  Phys.\ Rev.\ D \textbf{71}, 065008 (2005) [hep-th/0412320].

   \bibitem{VAKostelecky6}
   Q.G.~Bailey and V.A.~Kosteleck\'{y},
  {\em Signals for Lorentz violation in post-Newtonian gravity}, Phys. Rev. D \textbf{74}, 045001 (2006) [gr-qc/0603030].


 \bibitem{RBluhm}
   R.~Bluhm, N.L.~Gagne, R.~Potting and A.~Vrublevskis,
  {\em Constraints and stability in vector theories with spontaneous Lorentz violation},
   Phys. Rev. D \textbf{77}, 125007 (2008) [arXiv:0802.4071 [hep-th]].

 \bibitem{RCasana}
    R.~Casana, A.~Cavalcante, F.P.~Poulis and E.B.~Santos,
  {\em Exact Schwarzschild-like solution in a Einstein-bumblebee gravity model},
   Phys. Rev. D \textbf{97}, 104001 (2018) [arXiv:1711.02273 [gr-qc]].

\bibitem{DAGomes}
   D.A.~Gomes, R.V.~Maluf and C.A.S.~Almeida,
  {\em Thermodynamics of Schwarzschild-like black holes in modified gravity models},
   Annals Phys. \textbf{418}, 168198 (2020) [arXiv:1811.08503 [gr-qc]].


\bibitem{SKanzi}
  S.~Kanzi and $\dot{\text{I}}$.~Sakall{\i},
  {\em GUP Modified Hawking Radiation in Bumblebee Gravity},
  Nucl. Phys. B \textbf{946}, 114703 (2019) [arXiv:1905.00477 [hep-th]].


\bibitem{CDing1}
  C.~Ding, C.~Liu, R.~Casana and A.~Cavalcante,
  {\em Exact Kerr-like solution and its shadow in a gravity model with spontaneous Lorentz symmetry breaking},
  Eur. Phys. J. C \textbf{80}, 178 (2020) [arXiv:1910.02674 [gr-qc]].


 \bibitem{ZLi}
    Z.~Li and A.~\"{o}vg\"{u}n,
  {\em Finite-distance gravitational deflection of massive particles by a Kerr-like black hole in the Einstein-bumblebee gravity model},
   Phys. Rev. D \textbf{101}, 024040 (2020) [arXiv:2001.02074 [gr-qc]].

\bibitem{AAli}
  A.~Ali and K.~Saifullah,
 {\em Lorentz symmetry violating BTZ black holes in massive gravity}, (2020), arXiv:2004.02005 [gr-qc].

  \bibitem{SChen}
     S.~Chen, M.~Wang and J.~Jing,
  {\em Polarization effects in Kerr black hole shadow due to the coupling between photon and Einstein-bumblebee field},
   JHEP \textbf{2007}, 054 (2020) [arXiv:2004.08857 [gr-qc]].

\bibitem{RVMaluf}
  R.V.~Maluf and J.C.S.~Neves,
  {\em Black holes with a cosmological constant in bumblebee gravity},
  Phys. Rev. D \textbf{103}, 044002 (2021)
  [arXiv:2011.12841 [gr-qc]].

 \bibitem{SKJha1}
  S.K.~Jha and A.~Rahaman,
  {\em Bumblebee gravity with a Kerr-Sen-like solution and its Shadow}, Eur. Phys. J. C \textbf{81}, 345 (2021) [arXiv:2011.14916 [gr-qc]].

\bibitem{CDing2}
  C.~Ding, X.~Chen and X.~Fu,
 {\em Einstein-Gauss-Bonnet gravity coupled to bumblebee field in four dimensional spacetime}, (2021),
  arXiv:2102.13335 [gr-qc].

\bibitem{Carvalho}
  I.D.D.~Carvalho, G.~Alencar, W.M.~Mendes and R.R.~Landim,
  {\em The gravitational bending angle by static and spherically symmetric black holes in bumblebee gravity}, (2021), arXiv:2103.03845 [gr-qc].

 \bibitem{SKJha2}
  S.K.~Jha, S.~Aziz and A.~Rahaman,
  {\em Optical Properties of Lorentz violating Kerr-Sen-like spacetime in the presence of plasma}, (2021), arXiv:2103.17021 [gr-qc].

\bibitem{BCarter}
  B.~Carter,
  {\em Global structure of the Kerr family of gravitational fields},
  Phys. Rev. \textbf{174}, 1559 (1968).


\bibitem{JLSynge2}
 J.L.~Synge,
 {\em Relativity: the General Theory}, North Holland, Amsterdam (1960).

\bibitem{RCraigWalker}
  R.~Craig Walker, P.E.~Hardee, F.B.~Davies, C.~Ly and W.~Junor,
  {\em The Structure and Dynamics of the Subparsec Jet in M87 Based on 50 VLBA Observations over 17 Years at 43 GHz},
  Astrophys. J. \textbf{855}, 128 (2018) [arXiv:1802.06166 [astro-ph.HE]].

\bibitem{CBambi}
  C.~Bambi, K.~Freese, S.~Vagnozzi and L.~Visinelli,
  {\em Testing the rotational nature of the supermassive object M87* from the circularity and size of its first image},
  Phys. Rev. D \textbf{100}, 044057 (2019) [arXiv:1904.12983 [gr-qc]].


\end{thebibliography}
\end{document}